\documentclass[fleqn,usenatbib,useAMS]{mnras}
\usepackage[T1]{fontenc}
\usepackage{ae,aecompl}
\usepackage{threeparttable}
\usepackage{graphicx}	% Including figure files
\usepackage{amsmath}	% Advanced maths commands
\usepackage{amssymb}	% Extra maths symbols

\title[An Optical Analysis of A3888]{An Optical Analysis of the Merging Cluster Abell 3888}

\author[Shakouri, Johnston-Hollitt \& Dehghan]{
S. Shakouri,$^{1}$\thanks{E-mail: s.shakouri@vuw.ac.nz}
M. Johnston-Hollitt,$^{1}$
S. Dehghan,$^{1}$
\\
$^{1}$School of Chemical and Physical Science, Victoria University of Wellington, P.O.Box 600, Wellington 6140, New Zealand
}

\date{Accepted 2016 February 11. Received 2016 February 10; in original form 2016 January 08.}
\pubyear{2016}

\begin{document}
\label{firstpage}
\pagerange{\pageref{firstpage}--\pageref{lastpage}}
\maketitle

\begin{abstract}

In this paper we present new AAOmega spectroscopy of 254 galaxies within a 30$\arcmin$ 
radius around Abell 3888. We combine these data with the existing redshifts measured in a one 
degree radius around the cluster and performed a substructure analysis. We confirm 71 member galaxies within the core of A3888 
and determine a new average redshift and velocity dispersion for the cluster of 0.1535$\pm$0.0009 and 1181$\pm$197 km/s, 
respectively. The cluster is elongated along an East-West axis and we find the core is bimodal along this axis with two 
sub-groups of 26 and 41 members detected. Our results suggest that A3888 is a merging system putting to rest the previous conjecture 
about the morphological status of the cluster derived from X-ray observations. In addition to the results on 
A3888 we also present six newly detected galaxy over-densities in the field, three of which we classify as new galaxy clusters. 

\end{abstract}


%%%%%%%%%%%%%%%%%%%%%%%%%%%%%%%%%%%%%%%%%%%%%%%%%%

%%%%%%%%%%%%%%%%% BODY OF PAPER %%%%%%%%%%%%%%%%%%

\section{Introduction}
\label{sec:sec1}
In the last few decades, with the advent of multi-object spectrographs, astronomers have been capable of investigating the redshift information of galaxies over large areas
of the sky. Such wide-field analyses of spectroscopic surveys are thus an important tool to 
unravel the history and the underlying physics that plays a major role in the large scale structure formation in our Universe. 
Recent large redshift survey campaigns e.g. the Six-degree-Field Galaxy Survey (6dFGS, \citealt{2009MNRAS.399..683J}), 
the Two-degree-Field Galaxy Redshift Survey (2dFGRS, \citealt{1999lssu.conf..105C}) and the Sloan Digital Sky Survey (SDSS, \citealt{2009ApJS..182..543A}) have revealed that 
the visible matter creates web-like filamentary structures \citep{1985ApJ...299....5B, 2005MNRAS.364.1387P, 2014MNRAS.438.3465T} of the putative ``cosmic web'' on scales larger than a few megaparsecs. 
There are also megaparsec scale voids which are regions that harbour only small numbers of galaxies and fill the space between the filaments 
\citep{2004ApJ...607..751H, 2011IJMPS...1...41V, 2012ApJ...761...44S, 2014MNRAS.442.3127S, 2014AJ....147...52D}.

In conjunction with the large redshift surveys, large N-body simulations such as the Bolshoi simulation \citep{2011ApJ...740..102K} and the Millennium Run Observatory (MRObs) project \citep{2013MNRAS.428..778O}
are used to understand the underlying physics of the structure formation in the Universe. The N-body simulations \citep{1991ComPh...5..164B, 1997ApJS..109..307R, 2001NewA....6...79S, 2002MNRAS.336..409B} have confirmed the findings of the large redshift surveys; 
galaxies in the Universe form in large scales features known as filaments, sheets, superclusters
and cluster of galaxies \citep{1974ApJ...187..425P, 1989Sci...246..897G, 2005ApJ...624..463G}. 
It is known that the interconnecting nodes of filaments are the most common places to host galaxy clusters. 

One of the remarkable results from the aforementioned redshift censuses is the confirmation of the ``hierarchical model of cosmology'' \citep{1980lssu.book.....P, 2006Natur.440.1137S} 
which is currently largely accepted as the best model to explain large structure formation in the Universe. 
In this model, smaller units e.g. galaxies and galaxy groups, come together over a long period of time and merge to form larger units such as galaxy clusters 
\citep{2002AJ....123.1216R, 2005Natur.435..629S, 2011A&A...527A..78F, 2011ApJ...728...27O, 2015MNRAS.452.1617H}. 
According to this model of cosmology, merging of galaxy clusters is a very common phenomena and might happen multiple times during the life-time of galaxy clusters. 
It is believed that cluster merging is a key parameter in formation and evolution of galaxy clusters \citep{1997AJ....113..492C, 1998ApJ...493...62R, 2012ARA&A..50..353K}. Moreover, merging has a significant impact on cluster characteristics such as 
velocity dispersion \citep{1996MNRAS.279..349D, 2013A&A...556A..74R}, temperature \citep{1994Natur.372..439B, 2010A&ARv..18..127B}, mass \citep{2002ApJ...577..579R} and properties of the constituent galaxies \citep{2008MNRAS.390..289J, 2013MNRAS.433..825L, 2013MNRAS.429.1827P, 2016MNRAS.456..300O}. In addition, the hot, X-ray emitting gas which fills a vast region in the central part of the clusters (also known as 
Intra-Cluster Medium, ICM), is 
affected dramatically in the merging process via the powerful shocks \citep{2002ApJ...567L..27M, 2005ApJ...627..733M} and turbulence 
\citep{2006MNRAS.366.1437S, 2007MNRAS.378..245B, 2011MNRAS.410..127B} resulting from the cluster merger.

Merging of clusters often generates 
 clumps of galaxies within the cluster volume and this changes in galaxy volumetric density is known as 
 ``substructure'' \citep{1977Ap&SS..46..327W, 1982PASP...94..421G}.
Detection of such structure is an important tool to infer the cluster's dynamics and morphology.
The presence of substructure indicates
that the cluster is dynamically young and might be a signpost of an ongoing or past merger event. 

The dynamical status of a cluster is
determined via the optical spectroscopy of member galaxies and X-ray observations of the ICM. The brightness distribution of the X-ray 
images in merging clusters are asymmetric and often exhibit substructures \citep{1984ApJ...276...38J, 1986RvMP...58....1S}. However, it should be noted that the X-ray observations of a cluster are merely useful to 
detect merging events in or close to the plane of sky and  
are not sufficiently sensitive to the merging activities occur along our line-of-sight (LOS) 
\citep{2008A&A...479....1J, 2009ApJ...693..901O}. On the other hand, 
optical observations of clusters carry information about the merging not only happening in our LOS but in most of the cases with various merging axes.
Thus, the combination of the optical and X-ray substructure analyses of clusters is the most robust method to detect merging and its primary axis in clusters \citep{2013ApJ...772..104O, 2013MNRAS.432..243P, 2016MNRAS.456.2829G}.

In this paper, we present the results of new spectroscopic observations of the galaxy cluster Abell 3888 (A3888) and undertake an optical 
substructure analysis of the cluster. Identification of substructures in the galaxy clusters can be performed in one, 
two or three dimensions using either the radial velocity, spatial information or more comprehensively taking into account the 
spatial and velocity information together. Thus there is a large variety of substructure tests commonly used in analysis of clusters 
e.g. \citep{1979ASA.74..708, 1988MNRAS.230..161F, 1994AJ....108.2348A, 1996ApJ...458..435C, 1988AJ.....95..985D}
However, detailed comparison of statistical substructure tests has shown that they do not have the same sensitivity 
to structures in all clusters, mostly due to the physical orientation of the cluster, as a result no single 
test is adequate to properly characterise the dynamical status of a cluster \citep{1996ApJS..104....1P}. 
It is thus typical to perform a number of tests on spectroscopic data to adequately probe the dynamics 
of a system and this is the approach we have adopted here. The paper is arranged as follows: Section 
2 discusses the known properties of A3888, Sections 3 and 4 presents the AAOmega observations and data reduction. 
Section 5 presents the new redshifts determined here, while Sections 6 to 10 discuss the spectroscopic completeness, 
the cluster membership and substructure analysis of the cluster. Section 11 discusses newly detected fore and background 
groups and clusters and Section 12 presents the discussion and conclusions. In the following we assume the $\Lambda$CDM cosmology
with H$_{0}$=67 km s$^{-1}$Mpc$^{-1}$, $\Omega_{m}$=0.32 and $\Omega_{\Lambda}$=0.68
\citep{2014A&A...571A..16P}. At the redshift of A3888 1 Mpc = 6$\arcmin$.

\section{Properties of A3888}
\label{sec:morph}
This cluster has been previously studied at different wavelengths in large surveys such as 
the Representative XMM-Newton Cluster Structure Survey (REXCESS, \citealt{2007A&A...469..363B}), 
the Las Campanas AAT Rich Cluster Survey (LARCS, \citealt{2006MNRAS.366..645P}), 
the Local Cluster Substructure Survey (LoCuSS, \citealt{2010ApJ...716.1118P}) and is 
presented in the Planck early results VIII \citep{2011A&A...536A...8P}. Its X-ray properties 
have been reported in several publications, for instance, \citet{2009A&A...498..361P} 
measured the X-ray luminosity of the cluster and pointed out that A3888 is an X-ray luminous cluster with 
 L$_{x,500}$ = 6.363$ \times$ 10$^{44}$erg s$^{-1}$ [0.1--2.4 keV] which indicates this is a massive cluster 
 with M$_{500}$= 7.36 $\times 10^{14} M_{\sun}$ \citep{2010A&A...511A..85P}.\footnote{It should be noted that there is a large discrepancy between values of the reported X-ray luminosity of this cluster in the literature.
 \citet{1996MNRAS.281..799E} stated the X-ray luminosity to be 14.5$ \times$ 10$^{44}$erg s$^{-1}$ measured in the 0.1--2.4 keV band.  
 Later on, \citet{2002ApJ...567..716R} reported L$_{x}$ = 10.51$ \times$ 10$^{44}$erg s$^{-1}$ [0.1--2.4 keV] and finally 
 \citet{2009A&A...498..361P} lowered it further to L$_{x,500}$ = 6.363$ \times$ 10$^{44}$erg s$^{-1}$. 
 This discrepancy could be caused by measuring the X-ray flux in different apertures and 
 slight changes due to using different cosmological parameters. For the purposes of this work we choose to adopt 
the \citet{2009A&A...498..361P} value.}
 
Meanwhile, inconclusive X-ray sub-structure analyses of A3888 have been carried out in multiple papers 
\citep{2009A&A...498..361P, 2010A&A...514A..32B, 2012A&A...548A..59C} variously reporting the cluster to 
be either relaxed or disturbed based on the centroid shift and/or third order ratio values. Most recently \citet{2013A&A...549A..19W} 
used a new morphology estimator to identify disturbed or relaxed clusters finding that A3888 was an 
intermediate cluster in terms of substructure with some local asymmetries confined 
to the core but with the global properties consistent with a relaxed system (Wie{\ss}mann, private communication).

Additionally, A3888 has no central cD galaxy \citep{2009AAS...21344823H}, but rather is comprised of three brightest cluster galaxies (BCGs), the middle of which is located 112 kpc away from the X-ray peak
position \citep{2010ApJ...713.1037H}. The lack of a cD galaxy and an offset of the BCG from the X-ray peak of more than 42 kpc is often indicative of dynamical interactions in a young cluster \citep{2012MNRAS.420.2120M}. 
\citet{2006AJ....131..168K} scrutinised the intra-cluster light profile (ICL) of A3888 and found that the ICL profile of this cluster has a 
double exponential function which suggests that there is 
ongoing dynamical activity in the centre of this cluster. They pointed out the flux (within a certain aperture) 
of the ICL is less that the predicted value for this massive system and 
this finding indicates that A3888 is a dynamically young cluster.

With the aim of studying the dynamics of the cluster from an optical perspective,
we searched the NASA Extragalactic Database (NED) in a one degree radius around A3888 for available redshifts.
There were 788 reliable spectroscopic redshifts available in the literature
primarily from the LARCS project \citep{2006MNRAS.366..645P}. 
The redshift of the cluster is reported to be 0.1528${\pm0.0003}$ based
on 201 candidate member galaxies up to 16 Mpc away from the cluster
centre \citep{2006MNRAS.366..645P}. Despite the 201 possible members, there were only 72 
galaxies with reliable redshifts in a 1 Mpc radius around the cluster. Due to the lack of 
detailed optical substructure analysis in the literature and the small number of available redshifts in the core of A3888,
we opted for new optical observations of A3888 with the AAOmega spectrograph \citep{2006SPIE.6269E..0GS} to perform an 
optical analysis with the aim of achieving clear insight into the morphology of the cluster.

\section{AAOmega Observation of A3888}\label{sec:sec2}
We examined the Digitised Sky Survey 
(DSS) blue image and found that roughly two thirds of the galaxies up to the cluster's 
Abell radius ($\frac{1.72\arcmin}{z_{cl}}$ = 12$\arcmin$) 
did not have measured redshifts and the aforementioned redshift sample was not
isotropic in terms of coverage particularly in a 1 Mpc radius around the cluster. 
This anisotropy is not due to biased sample selection in the previous observations but rather is a result of 
the limitations of multi-object spectroscopy in sampling the dense cores of galaxy clusters.
Careful examination of the SuperCOSMOS catalogue, which has an astrometric accuracy of 0.3$\arcsec$ and completeness
of 90$\%$ up to m$_{b}$=21 \citep{2001MNRAS.326.1279H}, suggested that the majority of the galaxies without
measured redshift fall between blue magnitudes of 18 and 21.5.

AAOmega is a fibre-fed spectrograph which provides multi-object spectroscopy mounted at the Anglo Australian Telescope
(AAT). The spectrograph has 400 fibres covering a 
two-degree field when projected on sky making it an ideal instrument for examining the optical substructure in nearby southern clusters. 
Each fibre has a 2$\arcsec$ diameter when it
is projected on the sky and all fibres are placed by a robot. The accuracy
of locating the fibre on the target is about 0.3$''$ on the projected
sky which is well matched to the SuperCOSMOS astrometry. During observations, eight guide star fibres are used for accurate telescope
positioning. AAOmega has a dual beam which allows the observations of a wide range of
wavelengths. 

Since the number of targets in the AAOmega spectrograph configuration input file should
not exceed 800, we first selected all the galaxies with 17.5 $\le m_{b}
\le$ 21.5. In the next step, we excluded all the objects with available redshifts (in the NED) and  
then we sorted the remaining galaxies based on their cluster-centric radius;  
as a result, there were 790 galaxies up to 30$'$ radius from the centre of the cluster.
All of the 790 galaxies were selected as our science targets.

Although there are only 400 optical fibres available on the AAOmega spectrograph, 
for the sake of homogeneous observations it is recommended to include
between 500 and 800 science targets in the configuration file. This is particularly important when
the centre of the field is over-dense as the physical size of the fibre buttons prevents very close galaxies from simultaneously being observed.
This implies one should have a larger number of targets and optimise accordingly using the field configuration software produced by the observatory.
For the 790 galaxies included in our configuration file, the highest priority was given to the targets which were inside the Abell radius (12$\arcmin$). Objects between 12$\arcmin$ to 25$\arcmin$ 
(4 Mpc) had the next priority and the remaining targets between 25$\arcmin$ to 30$\arcmin$were given the minimum priority in the configuration file. This thus produced a field configuration which was biased towards galaxies within a 25$\arcmin$ radius of the cluster. There was no priority assigned to objects based on their magnitude with equal priority given to all objects up to a blue magnitude of 21.5.

Previous spectroscopy of A3888 was primarily carried
out in the LARCS project in which the targets were selected such that galaxies within certain radii of the cluster 
were prioritised based on their absolute magnitude such that for magnitudes R = 16.5 and M$_{v}$ = -21.8 the 
highest priority was given to objects within 30$\arcmin$ for which R < R(M$_{v}$ + 1)], next were objects 
within that same radius which had  R(M$_{v}$ + 1) < R < R(M$_{v}$ + 3)] and finally objects more that 
30 $\arcmin$ in radius with R < R(M$_{v}$ + 1)]. Thus whereas the LARCS selection was based on cluster-centric distance and absolute magnitude, 
our selection was made based on the cluster-centric distance only for objects up to a blue magnitude of 21.5.

Our new observations of the cluster A3888 were taken on the 9th and 10th of May 2013
with the AAT in service mode. A total integration 
time of three hours was observed in half an hour blocks so as to reduce cosmic ray contamination. The typical seeing was
1.5$\arcsec$ to 2$\arcsec$ for both nights.
A total of 337 optical fibres were allocated to the science targets. 
There were 8 and 25 fibres allocated to guide stars and sky positions,
respectively and 30 
fibres remained un-allocated since some of them were broken or they had been detached during the fibre un-crossing procedure.
All the sky positions were visually inspected using the DSS blue image to ensure
that there were no optical sources within a 5$\arcsec$ radius 
around each of the sky positions. 
We used a low to medium spectral resolution where the blue arm has a 580V grating
at the central wavelength of 4800{\AA} and the red arm has a 385R grating with the
central wavelength 7250{\AA}. 
This set of gratings provide a spectral resolutions of 3.5{\AA}
and 5.3{\AA} at 4800{\AA} and 7250{\AA}, respectively. The combined wavelength coverage resultant from both arms
was 3700--8800{\AA}. This wavelength coverage allows detection of emission lines such as H$\alpha$, H$\beta$, H$\delta$, H$\gamma$, NII, SII , [OI], [OII] and 
[OIII], and absorption lines such as H, K, G, Na I and Mg up to redshift of $\sim$ 0.5 and it is well matched to the redshifts of nearby clusters.

\section{Data Reduction and Redshift Determination}
\label{sec:redshift-detemine}
The observatory staff performed the observations collecting the required calibration files including bias, dark, arc and flat field frames.
The data were automatically reduced using the AAO data reduction
pipeline software ``2dfdr'' \citep{2001MNRAS.328.1039C}. We briefly outline the calibration procedure here.
First, the bias image is generated during the reading process of the CCD whilst it is not exposed to light over a zero second interval to measure the inherent 
noise generated by the CCD amplifiers. Next, a
dark frame is taken when the CCD is not exposed to light (shutter closed) but the exposure time is 
identical to the exposure time of the object frames (science image). Dark frames are usually used to measure
the thermal noise across the CCD array. Flat field exposures are then taken and used to correct the
photon capture variation of the pixels across the CCD. Flat fields are used to illuminate a
CCD in a long exposure to generate a
calibration image with a very high signal-to-noise ratio. Finally, CuAr, helium and FeAr arc lamp images are usually used to generate
spectral lines to calibrate the wavelength of the object frames. In addition to above instrumental calibration issues, the sky itself also needs to be accounted for.
Cosmic rays hit the CCDs during an exposure producing bright patches which are accounted for with a sigma-clipping algorithm within 2dfdr.
Additionally, water vapour, O$_{3}$ and O$_{2}$ in the Earth's atmosphere can cause significant imprints in the target's spectra by generating false absorption features 
which are usually referred to as
Telluric absorption spectral lines. These absorption lines are dominant in near the Infra-red (IR)
and visible regions of the spectra. There are also contaminating emission lines known as 
sky lines that are mainly caused by NaI and OH in the atmosphere. 
The 2dfdr software calibrates and subtracts night sky lines (atmospheric emission and absorption
lines) automatically for each multi-object frame and for each arm separately. 
After reducing each frame, all of the frames from each spectrograph were combined. 
Afterwards, the coadded frames for the blue and red arms are
spliced together. 

This final reduced and spliced image can be used as an input file to any spectroscopic
redshift fitting software. In this work, we used the ``RUNZ'' code to extract the redshifts from our spectra.\footnote{ 
RUNZ was originally developed by Dr Will Sutherland  
with the aim of using it in the two degree field Galaxy Redshift Survey (2dfGRS, \citealt{2001MNRAS.328.1039C}). 
Currently, RUNZ is available and maintained by to A/Prof. Scott Croom who kindly provided the software to us.}

RUNZ measures two redshifts simultaneously employing two different methods:
\begin{itemize}
\item Emission line redshift: An individual redshift is measured by fitting a Gaussian to each
emission feature in the spectrum and the final redshift is assigned to an object by calculating the mean value of the 
variance weighted measured redshifts. 
\item Cross-correlation redshift: This redshift is assigned to each spectrum
by employing the standard cross-correlation technique \citep{1979AJ.....84.1511T}. Here,
each spectrum is cross-correlated to a set of stellar and galaxy spectral
templates.
The redshift with the highest peak of the correlation coefficient is
considered as the final cross-correlated redshift. The technique is primarily concerned with absorption features, and the success of this method depends strongly on the 
quality of the input templates.
\end{itemize}
The Sloan Digital Sky Survey data release 7 (SDSS DR7) galactic and stellar spectroscopic templates were used in order to determine the
cross-correlated redshifts. In addition to the estimated redshift, RUNZ also returns an automatically assigned redshift quality flag ``q'' ( ranging from 1--5) to 
describe the reliability of the estimated redshift. Redshifts with q$\le$2 are
considered unreliable, q = 3 is considered likely and redshifts with q$\ge$4 are very reliable. 
For the qualitative definitions of the automatic quality flags we refer the reader to \citep{2001MNRAS.328.1039C}. All of the spectra were visually
inspected to confirm the accuracy of the redshift measurement and the software-determined redshift quality flag. The
measured redshift (emission or absorption) with the higher quality flag was reported as the final redshift for each spectrum. 
There was no spectrum with the same quality flags for both the emission and the absorption measured redshifts.

\begin{figure}
\centering
 \hspace{1mm}
 \includegraphics[scale=0.34,clip,trim=0 15.5mm -2mm 0]{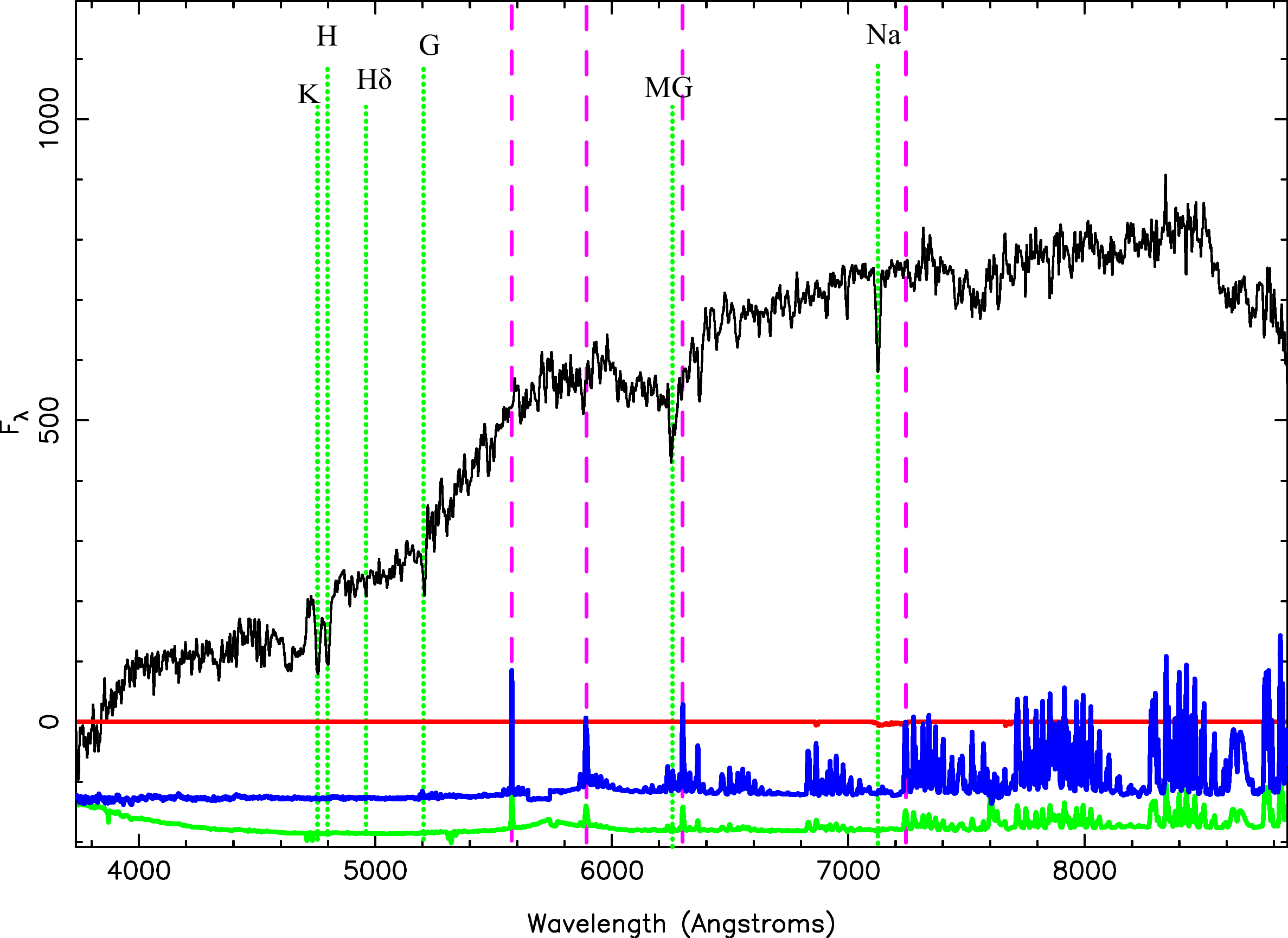}\\
  \hspace{0.3mm}
\includegraphics[scale=0.34,clip,trim=0mm 0 0mm 0mm]{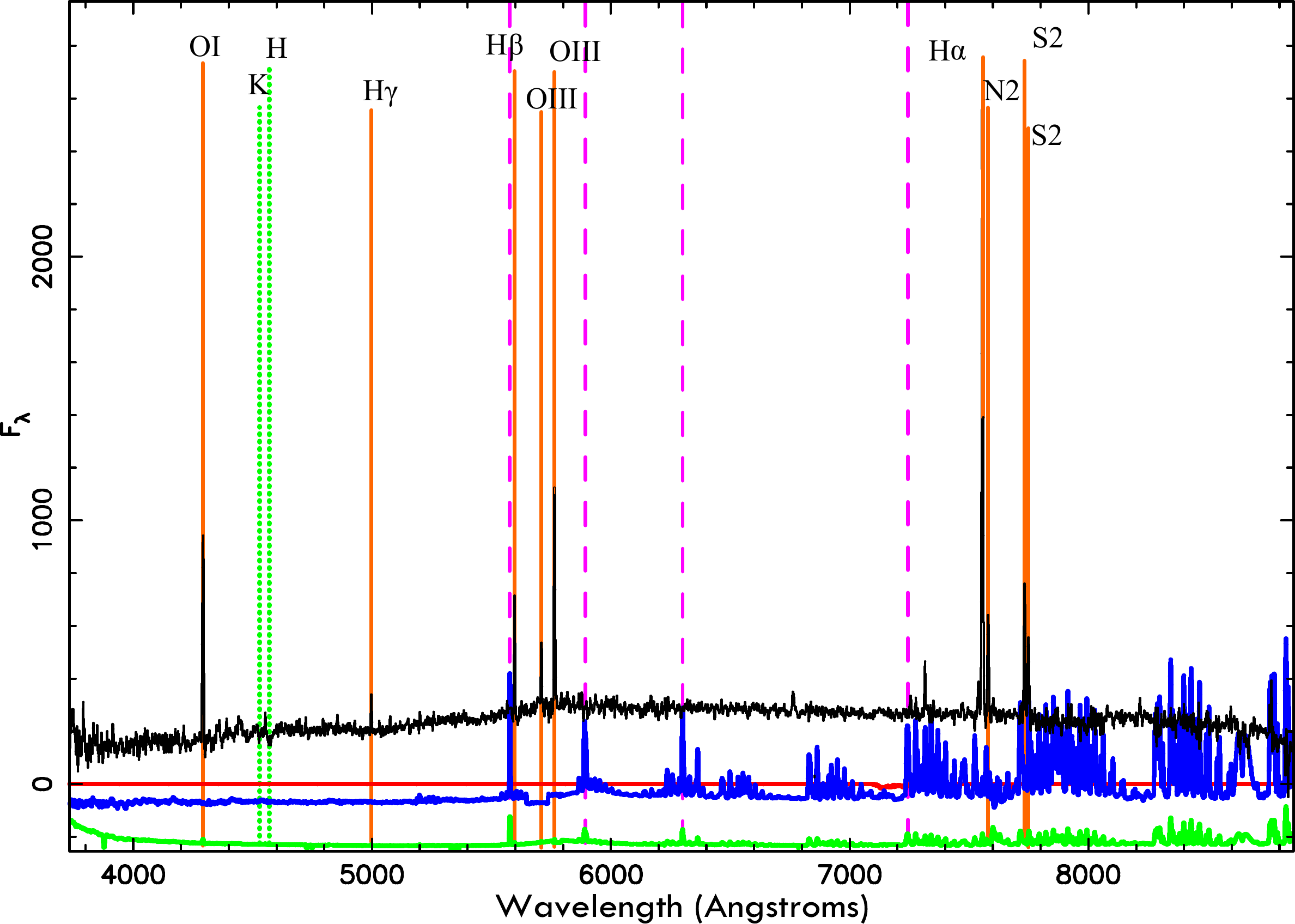}
\caption[Two examples of reduced spectra observed with the AAOmega]{ {Two examples of reduced spectra are shown. Top panel: An
absorption line galaxy in the field with redshift quality 5. 
 Bottom panel: An emission line galaxy in the field with redshift quality of 5.
 In each plot,
the galaxy spectrum (black line), residual spectrum after sky subtraction
(green or light grey spectra in the black and white version), 
sky spectrum (blue / dark grey spectra) 
and the Telluric absorption spectrum (red or grey horizontal line) are
superimposed. The emission (solid orange or medium grey vertical lines) and absorption (dotted green or light grey vertical lines) features at the
best fitted redshift are also shown. The most intensive night sky emission lines are demonstrated with dashed vertical magenta (medium grey) lines.}}
 \label{fig:spectra}
\end{figure}

In some cases for absorption redshifts, we
assigned a different redshift quality from the one which was assigned by the
RUNZ automatically. We manually assigned a quality of 4 to redshifts where both H and K absorption lines were present.
In the case that H and K and at least one of the other absorption line (Mg, Na,
G, H$\delta$) were present, a quality of 5 was assigned to the redshift.
If merely H or K and at least two of the other absorption lines were present a
quality of 4 was assigned. In the case that neither H or K were present 
if all the other absorption lines were present then the quality of 3 was given.
In the case that only one or two absorption lines (Mg, Na, G, H$\delta$)
were present then a quality of 2 was given to the redshift.
In the case that no absorption features were
detectable then a quality of 1 was assigned.
In Figure~\ref{fig:spectra}, examples of emission and absorption spectra and associated fitted lines from RUNZ
are shown.

%%%%%%%%%%%%%%%%%%%%%%%%%%%%%%%%
%%%%%%%%%%%%%%%%%%%%%%%%%%%%%%%%

\section{Redshift Extraction Results}
\label{sec:sec4}
Following the redshift extraction using RUNZ, we found that 41 objects of our 337 science targets were stars (88\%)
which is consistent with the uncertainty level of the SuperCOSMOS star/galaxy classification, that \citet{2001MNRAS.326.1295H} estimated to be $\sim$90\% accurate down to m$_{b}$ = 21.
A total of 42 measured redshifts from objects with fairly faint blue magnitudes (18.7 $< m_{b} <$ 20.5) were assigned 
q$\le$2 thus they were not included in our analysis. \citet{2001MNRAS.328.1039C}
pointed out that the successful redshift determination rate decreases significantly beyond $m_{b}$ = 19 and our detection success rate (87.5\%) 
is in agreement with their finding for the entire 2dF Galaxy Redshift Survey.
After the removal of the misclassified stars and non-detected galaxies the remaining 254 
targets observed here were classified as galaxies with reliable redshifts. We have supplemented
our redshift sample with available redshifts in the literature. We
searched the NED
up to one degree radius from the core of A3888 for reliable spectroscopic redshifts. We then co-added our
spectroscopic redshifts to the published redshifts to build a final redshift sample containing 1027
spectroscopic redshifts.

Table~\ref{tab:redshift-example} presents the details of the new spectroscopic redshifts obtained by AAOmega and presented in this work.
We note that despite the fact that we excluded the objects with available redshifts in the literature in our configuration file, ultimately we still had 15 galaxies in common between our AAOmega observations and the LARCS project.
We find that this was because the reported positions in the NED have more than a 3$\arcsec$ separation from the reported positions in SuperCOSMOS and thus, they were missed in
the cross matching process. The list of the common objects between our AAOmega observation and the LARCS are given in Table~\ref{tab:common}.

%%%%%%%%%%%%%%%%%%%%%%%%%%%%%
\begin{table*}
\caption{Details of the objects observed with the AAOmega. 
Columns 1 and 2 give the J2000 coordinates of the object, 
Columns 3 and 4 give the redshift and its uncertainty, 
Column 5 is the quality flag, Column 6 is the spectral lines detected, 
Columns 7 gives the spectral type and 
finally Column 8 is the blue magnitude from the SuperCOSMOS catalogue. 
This Table presents the first 10 lines and the full table of 254 galaxies will be available online.}
  \scriptsize{
  \begin{tabular}{l l c c c l c c}
\hline
\\[-2mm]
RA	&	Dec	&	z 	&	error 	&
Quality 	&Spectral lines 	&	Spectral type &Blue mag\\
\\[-2mm]
\hline
 22  32 02.41	&	  -37  52 50.48	&	0.19935	&	0.00016	&	5  & OII,H$\alpha$,N2,S1,S2,H$\beta$		& Em	&18.96	\\
 22  32 08.90	&	  -37  38 53.20	&	0.16385	&	0.00016	&	5  &H,K,H$\delta$,G,Na				& Abs	&19.46	\\
 22  32 11.88	&	  -37  43 11.60	&	0.20863	&	0.00013	&	5  &OII,H$\alpha$,N2,S1,S2,H$\beta$,OIII	& Em	&19.35	\\
 22  32 15.43	&	  -37  37 23.34	&	0.03880	&	0.00010	&	5  &OII,H$\alpha$,N2,S1,S2,H$\beta$,OIII	& Em	&17.57	\\
 22  32 16.22	&	  -37  52 27.08	&	0.07337	&	0.00010	&	5  &OII,H$\alpha$,N2,S1,S2,H$\beta$,OIII	& Em	&18.64	\\
 22  32 19.76	&	  -37  41 35.63	&	0.03838	&	0.00016	&	5  &OII,H$\alpha$,N2,S1,S2,H$\beta$,OIII	& Em	&20.07	\\
 22  32 20.90	&	  -37  42 53.93	&	0.38120	&	0.00011	&	5  &OII,H$\alpha$,N2,S1,S2,H$\beta$,OIII	& Em	&19.09	\\
 22  32 21.10	&	  -37  55 55.02	&	0.14667	&	0.00015	&	5  &OII,H$\alpha$,S1,S2,N2			& Em	&19.22	\\
 22  32 21.34	&	  -37  30 30.92	&	0.20933	&	0.00013	&	5  &OII,H$\alpha$,N2,S1,S2,H$\beta$,OIII	& Em	&18.97	\\
 22  32 21.54	&	  -37  51 09.72	&	0.03188	&	0.00017	&	5  &OII,H$\alpha$,N2,S1,S2,H$\beta$,OIII	&Em	&19.40\\
 \\[-2mm]
\hline
\end{tabular}
}
\label{tab:redshift-example}
\end{table*}
%%%%%%%%%%%%%%%%%%%%%%%%%%%

%%%%%%%%%%%%%%%%%%%%%%%%%%%
\begin{table}
\caption{List of common observed galaxies. All the common galaxies were also observed in the LARCS project.}
\scriptsize{
 \begin{tabular}{ccccc}
 \hline
 \\[-10pt]
 RA $_{(J2000)}$, Dec$_{(J2000)}$& z$_{AAOmega}$& z$_{err}$& z$_{LARCS}$& z$_{err}$\\
 \\[-10pt]
 \hline
 \\[-10pt]
22 32 29.7\hspace{2mm}-37 37 34&0.13942&0.00025&0.13973&0.00140\\
22 33 13.3\hspace{2mm}-37 46 54&0.13993&0.00027&0.13933&0.00260\\
22 33 19.5\hspace{2mm}-37 42 12&0.13133&0.00031&0.13182&0.00056\\
22 33 42.0\hspace{2mm}-37 45 45&0.15194&0.00022&0.15213&0.00050\\
22 33 54.4\hspace{2mm}-37 42 34&0.14702&0.00036&0.14696&0.00011\\
22 34 16.8\hspace{2mm}-37 49 13&0.15162&0.00033&0.15140&0.00007\\
22 34 22.5\hspace{2mm}-38 03 38&0.12511&0.00058&0.12546&0.00010\\
22 34 42.3\hspace{2mm}-37 22 40&0.20857&0.00058&0.20890&0.00110\\
22 34 45.2\hspace{2mm}-37 39 51&0.15590&0.00037&0.16223&0.00009\\
22 34 49.7\hspace{2mm}-37 29 52&0.14086&0.00016&0.14505&0.00010\\
22 34 49.9\hspace{2mm}-37 47 50&0.15250&0.00039&0.15282&0.00020\\
22 35 27.0\hspace{2mm}-37 25 42&0.12576&0.00005&0.12570&0.00096\\
22 35 45.3\hspace{2mm}-37 21 23&0.14065&0.00042&0.14491&0.00060\\
22 36 03.9\hspace{2mm}-37 39 09&0.20138&0.00016&0.20208&0.00120\\
22 36 19.5\hspace{2mm}-37 38 30&0.19754&0.00016&0.19740&0.00230\\

\\[-10pt]
\hline
\end{tabular}
}
\label{tab:common}
\end{table}
%%%%%%%%%%%%%%%%%%%%%%%%%%%

\begin{figure}
\centering
 \includegraphics[scale=0.65]{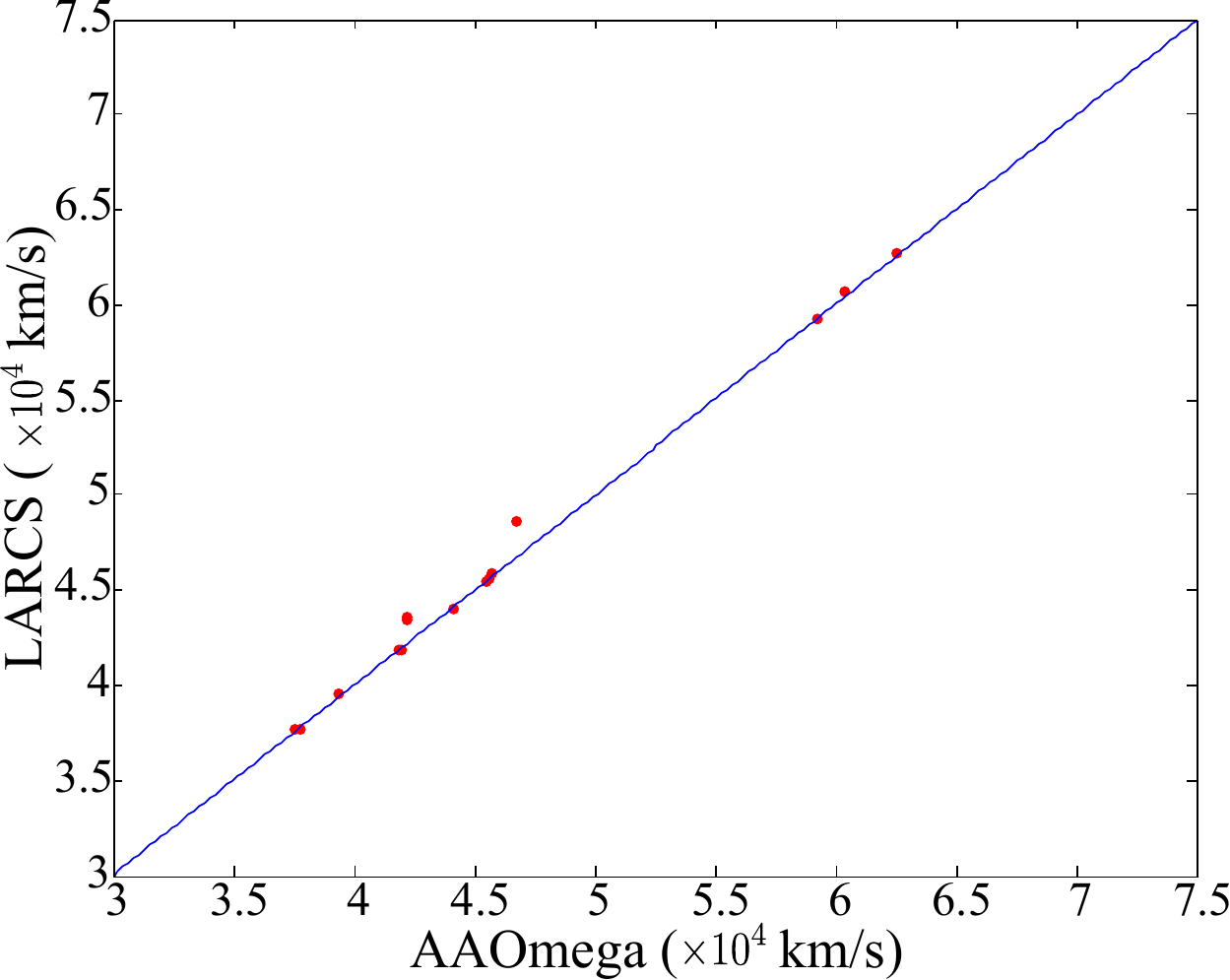}
 \caption{Common objects between the AAOmega observations and the literature, the error bars are not shown due to being
 smaller than the size of the dots on the plot (the error mean value is about 0.0003 in z or 90 km/s). The solid line is the one-to-one relation.}
 \label{fig:common}
\end{figure}

For cases where an object had
two redshift measurements, our redshift was chosen if
q$\geq$4 otherwise the redshift from the literature was used in our
analysis. For all such galaxies, the DSS blue image
was visually 
inspected to ensure that the cross-matching process did not conflate 
two separate objects and they were indeed true cross-matches.
Figure~\ref{fig:common} shows the consistency between our extracted redshifts and the redshifts from the literature (for our 15 common redshifts).
The plot shows that the majority of our common redshifts are in good agreement with the published redshifts.

%%%%%%%%%%%%%%%%%%%%%%%%%%%
%%%%%%%%%%%%%%%%%%%%%%%%%%%

\begin{figure}
\centering

 \includegraphics[scale=0.65,clip,trim=5mm 0 0 5mm s]{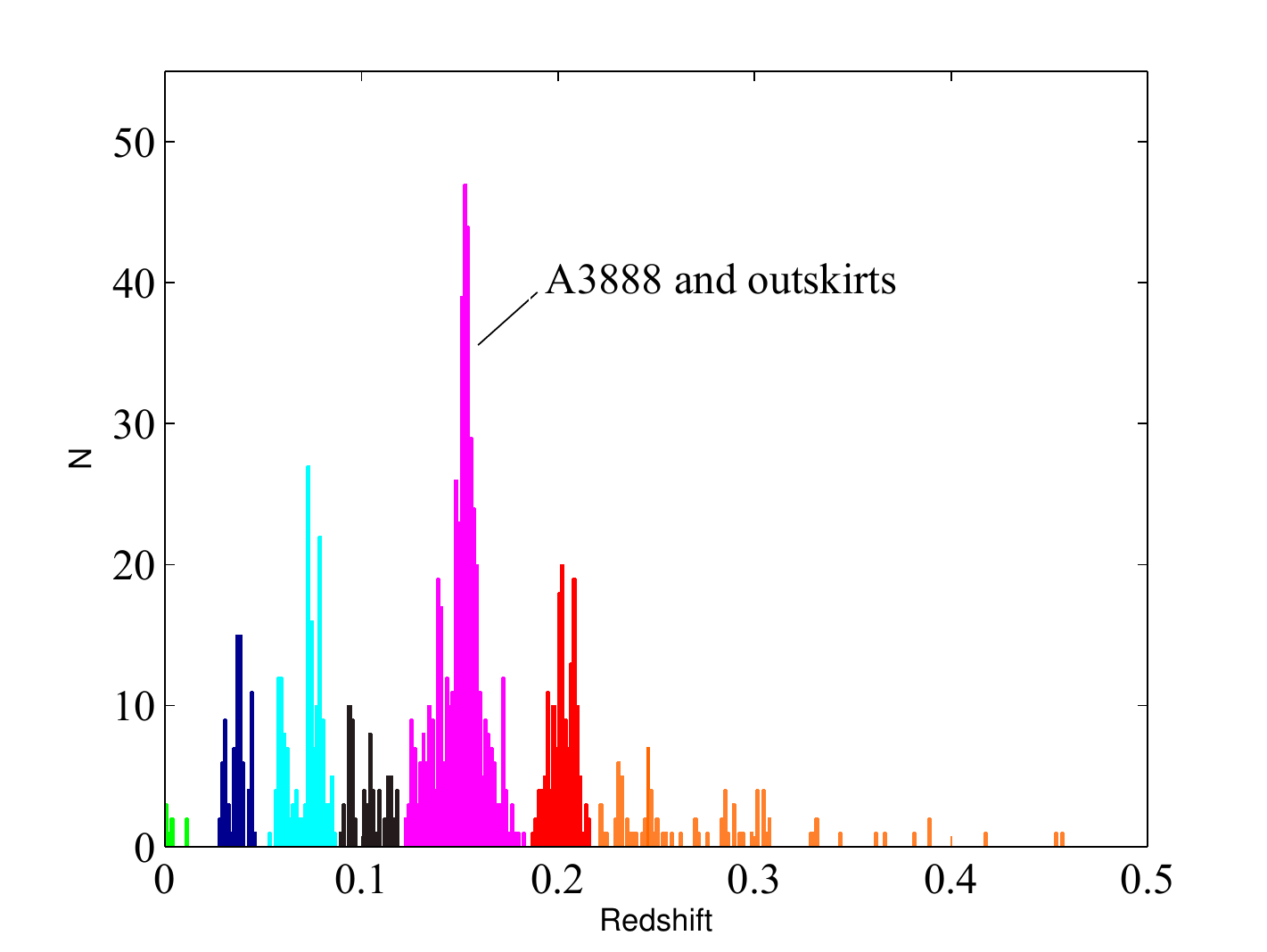} 
 \caption{The colour-coded redshift histogram of the combined redshifts (AAOmega + literature) in
a one degree radius around the core of A3888. 
According to this plot, there are seven distinguishable populations up to the redshift of 0.35. 
See Table~\ref{tab:zdist} for details. }
\label{fig:colorhist}
\end{figure}

A colour-coded histogram of the combined redshifts from the final catalogue in a one degree radius around
A3888 is shown in Figure~\ref{fig:colorhist}.
The redshift distribution clearly shows the
existence of seven distinct populations around A3888. The populations span the redshift range 0.0001--0.4580. 
The redshift range associated
with each population was determined based on conspicuous gaps in the redshift histogram.
In Table~\ref{tab:zdist}, the redshift range and the number of galaxies in each of identified velocity groups 
in the combined AAOmega and literature redshift sample are shown. 

%%%%%%%%%%%%%%%%%%%%%%%%%%%
%%%%%%%%%%%%%%%%%%%%%%%%%%%

%%%%%%%%%%%%%%%%%%%%%%%%%%%
\begin{table}
\centering
\caption[Details of the redshift in a one degree radius around A3888]
{ {Details of the velocity groups and number of galaxies in each redshift slice in a one degree radius around A3888.
The largest structure in velocity is clearly A3888 and its outskirts with 469 member galaxies and an extent of $\sim$10 Mpc in radius.}}
\begin{tabular}{ll}
\hline
\\[-10pt]
 Redshift Range & N$_{gal}$\\
 \\[-10pt]
\hline
\\[-10pt]
0.0001 -- 0.0108 &8\\
0.0284 -- 0.0475& 80\\ 
0.0530 -- 0.0875& 163 \\ 
0.0880 -- 0.1190& 67\\
0.1200 -- 0.1850&469\\
0.1850 -- 0.2220& 156\\
0.2230 -- 0.4580& 84\\
%0.2230 -- 0.2650& 43\\
%0.2670 -- 0.3080& 28\\
%0.3100--0.4580&13\\
 \\[-10pt]
\hline
\end{tabular}
\label{tab:zdist}
\end{table} 
%%%%%%%%%%%%%%%%%%%%%%%%%%%

\section{Spectroscopic Redshift Completeness}
\label{sec:sec5}
\begin{figure}
\centering
\includegraphics[scale=0.38]{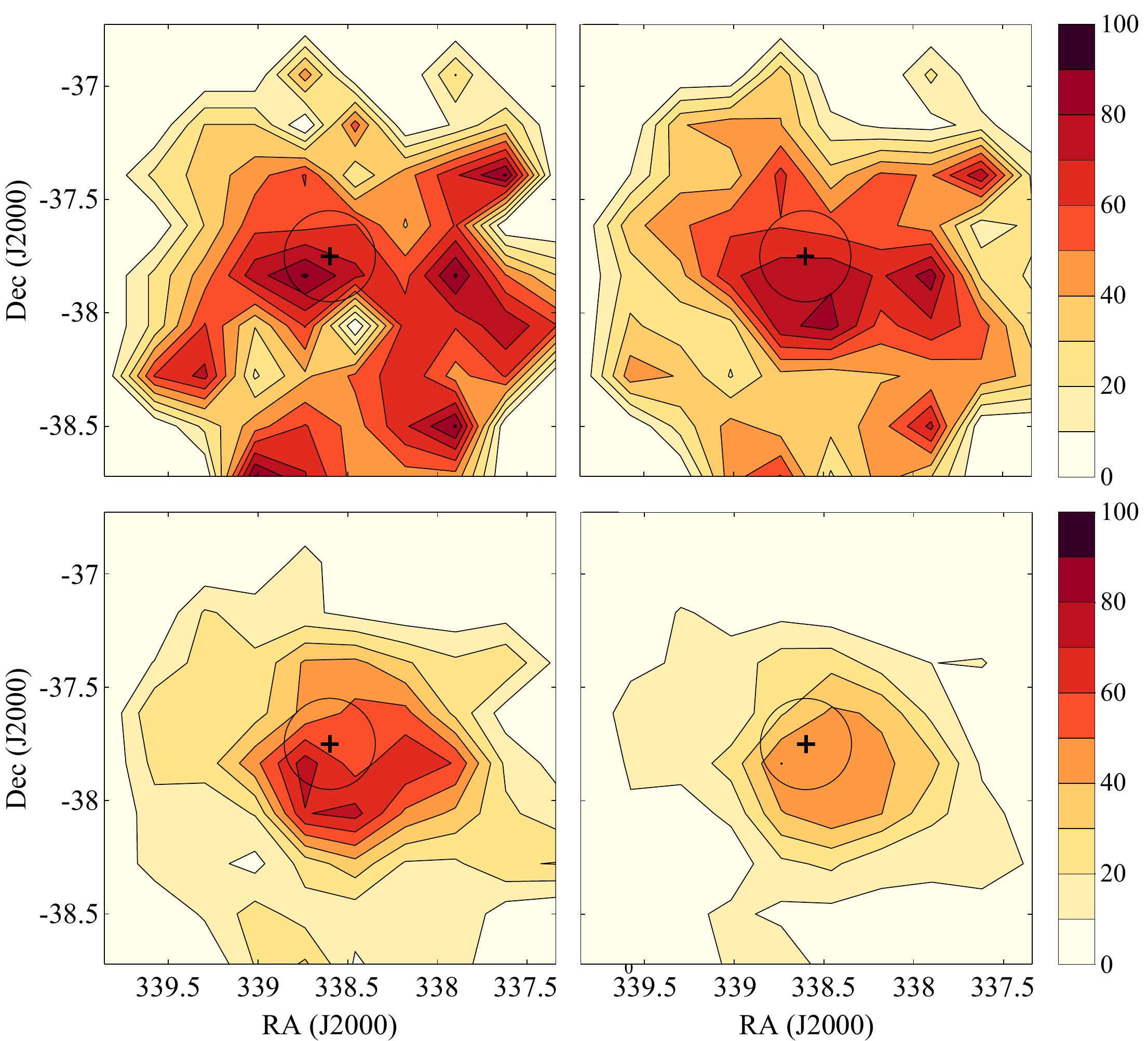}
  \caption{Spectroscopic completeness map of A3888. Top left panel: spectroscopic 
completeness in $m_{b} \leq$ 18. Top right panel: completeness in $ m_{b} \leq$ 19.
 Bottom left panel: completeness in  $m_{b} \leq$ 20. Bottom right panel: 
completeness in $m_{b} \leq$ 21. Black cross shows
the position of A3888 and the black circle shows a region with radius of 2 Mpc around A3888.}
 \label{fig:completenes-bin-2}
\end{figure}

Completeness of our redshift sample was determined by calculating
the ratio of galaxy number density in the combined redshift catalogue
(AAOmega+literature with the upper limit of m$_{b}$ = 21.32)
to the number density of
galaxies with available blue magnitudes in the SuperCOSMOS catalogue which is 90\% complete up to blue magnitude of 21. 
There were 937 galaxies out of 1027
galaxies in our redshift sample which had available blue magnitude in the SuperCOSMOS
catalogue. The final spectroscopic completeness was scaled by multiplying the
ratio of the number of redshifts with available blue magnitude to the total number of available
redshifts.
Figure~\ref{fig:completenes-bin-2} shows the cumulative spectroscopic completeness of our final
redshift sample as a function of blue magnitude. As it is expected, the spectroscopic completeness
decreases whilst the blue magnitude cut increases from top left to bottom right. In some regions the completeness increases, 
however this unexpected behaviour has been previously
seen in other works. For instance, \citet{2011ApJ...728...27O} presented the completeness of the spectroscopic observation of Abell 2744 
with the AAOmega and they have observed the same behaviour in some regions. This is suggestive of an anisotropic redshift coverage as a function of magnitude. 

%%%%%%%%%%%%%%%%%%%%%%%%%%%
%%%%%%%%%%%%%%%%%%%%%%%%%%%

\begin{figure*}
\centering
\begin{tabular}{cc}
\hspace{-0.6cm}
 \includegraphics[scale=0.68,clip,trim=0mm 0mm 0mm 0mm]{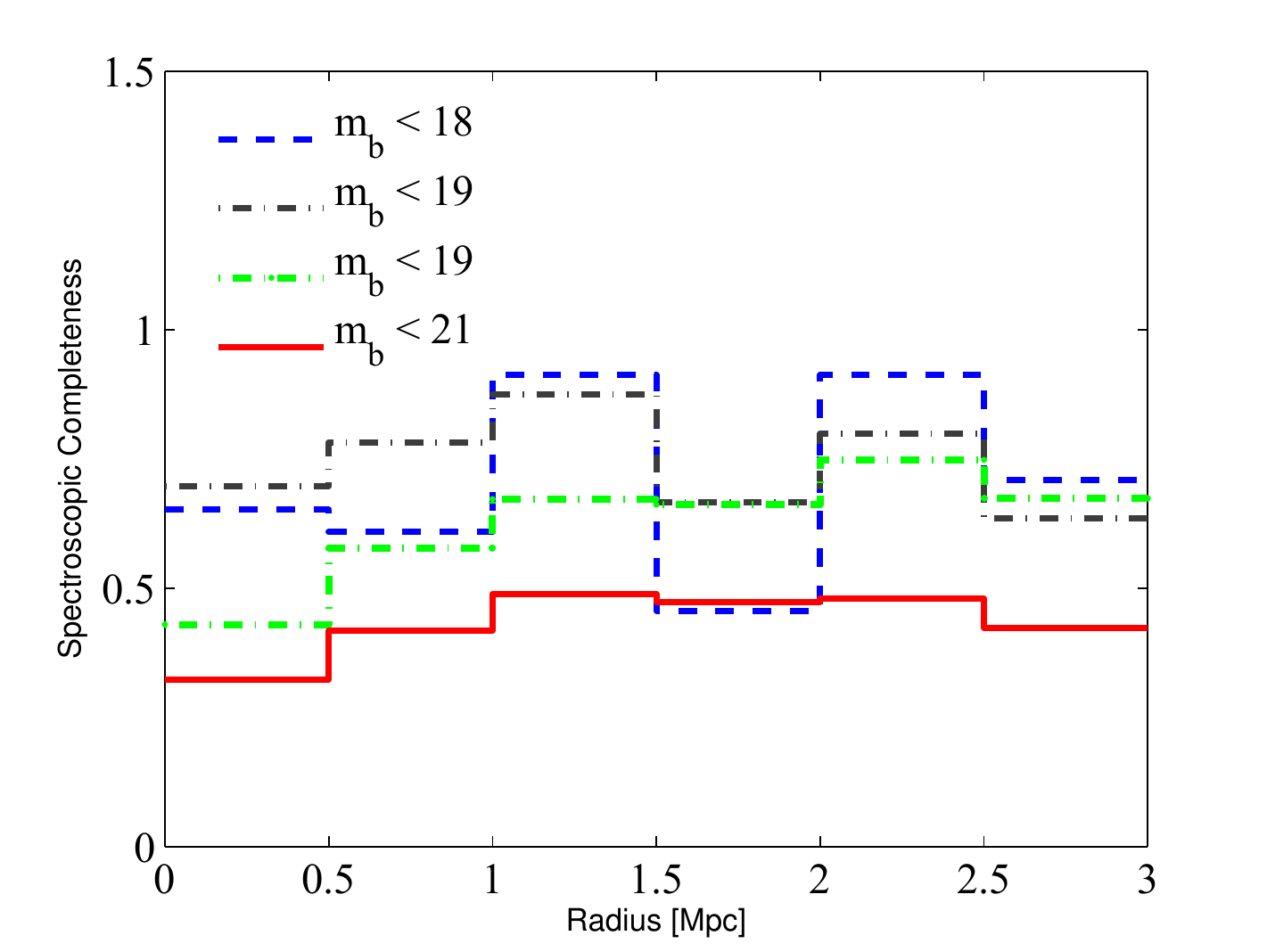}&
 \hspace{-0.8cm}
  \includegraphics[scale=0.68,clip,trim=10mm 0mm 0mm 0mm]{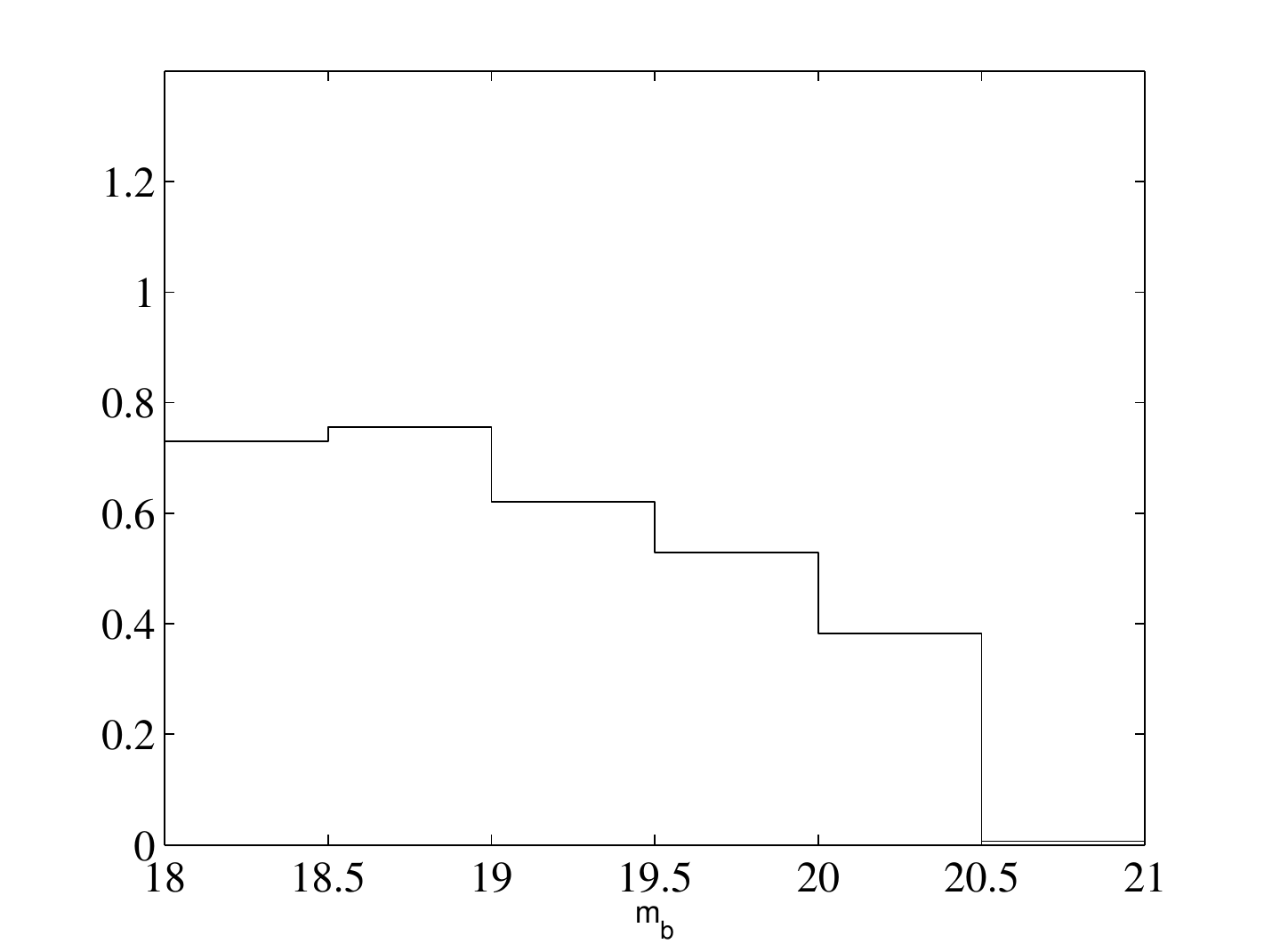}\\
  \end{tabular}
 \caption{Left panel: cumulative spectroscopic redshift completeness in radial bins of 500 kpc. Right panel: 
 spectroscopic redshift completeness as a function of magnitude within a 3 Mpc radius from the cluster core.}
\label{fig:completeness-3mpc}
\end{figure*}

Figure~\ref{fig:completenes-bin-2} also shows that 
the completeness decreases whilst moving radially outwards which is in agreement with giving a 
low priority to the outermost regions which are considered as the outskirts of A3888
in our AAOmega configuration procedure.
In Figure~\ref{fig:completeness-3mpc} the left plot shows the cumulative spectroscopic completeness as a function of
radius from the core of A3888. According to this plot the completeness is above 30\% in all radial bins for all magnitude ranges. 
The low completeness value in the first radial bin of 500 kpc from the cluster core 
indicates that the central region of the cluster was extremely over-dense and AAOmega was unable to cover the very dense cluster core due to the physical size of the fibre buttons 
3$\arcsec$, $\sim$2.77 kpc at the redshift of cluster)
and the minimum required fibre separation of 30$^{\arcsec}$.
The right plot in Figure~\ref{fig:completeness-3mpc} shows the redshift completeness in different magnitude bins within a 3 Mpc radius from A3888. 
This plot shows that the spectroscopic completeness
is at least 38\% within a 3 Mpc radius from the core of A3888 in all magnitude bins whilst the magnitude is brighter than m$_{b}$ = 20.5.

%%%%%%%%%%%%%%%%%%%%%%%%%%%
%%%%%%%%%%%%%%%%%%%%%%%%%%%

 \section{Cluster Membership}
\label{sec:sec6} 
Decades of optical observations of galaxy clusters have revealed that the majority of galaxy clusters host a single or 
multiple substructures allowing a broad classification as either dynamic or relaxed systems, respectively.  
The presence of
substructure in a cluster is likely indicative of ongoing dynamical interactions. 
Substructures may be formed through the infall of 
individual galaxies or galaxy groups into a relaxed cluster or during the merging of two or more entire galaxy clusters.  
According to theoretical studies and simulations of galaxy clusters, member galaxies of virialised clusters are expected to exhibit 
a well-behaved Gaussian velocity distribution \citep{2009ApJ...693..901O}. Therefore, 
any sign of non-Gaussian behaviour in the velocity distribution can be a signpost of certain conditions such as existence of fore/background  
interloper galaxies and more importantly, substructure in the
cluster \citep{1996ApJS..104....1P,2009ApJ...693..901O}. However, it
should be
noted that a well established single Gaussian in velocity distribution does not
necessarily indicate that the cluster is in dynamical equilibrium
\citep{2009ApJ...693..901O}. 
For instance, the cluster Abell 3667, which has been well studied in the past, is
known to be a highly disturbed cluster, however \citet{2008A&A...479....1J}
and \citet{2009ApJ...693..901O} 
pointed out that this cluster has a well described Gaussian velocity
distribution as measured along the line of sight, transverse to the known plane of the sky merger axis.

Additionally, interlopers may cause non-Gaussian behaviour in the velocity distribution which
can be mistaken as true substructure \citep{2009ApJ...693..901O}.
Hence, it is imperative to minimise fore/background galaxy contamination in the
cluster field. To achieve this, spatial and LOS peculiar velocity
distributions of galaxies are used solely or in a combined manner to determine the cluster
members before any further substructure analysis is under taken.

The interloper rejection and cluster member selection for A3888 was performed combining two methods: a Density-Based 
Spatial Clustering of Applications with Noise (DBSCAN) algorithm
\citep{eks96} and the caustic technique \citep{1997ApJ...481..633D}.

\subsection{DBSCAN Clustering Algorithm}
\label{subsec:subsec1}
The initial cut of the
foreground and background galaxies (see Section \ref{sec:sec4}) was carried out by identifying the
velocity gaps in the redshift histogram. Further outlier removal was performed by employing the
DBSCAN
method. This technique has been recently used in different fields of
Astrophysics to statistically aggregate data in complex, irregularly sampled datasets, 
for instance \citet{2013A&A...549A.138T} and \citet{2013PhRvD..88d3006C} adopted it in analysing the Fermi-LAT gamma ray
datasets. More recently \citet{2014AJ....147...52D} used this method to detect and allocate members to the galaxy groups and clusters in the Chandra Deep Field-South (CDFS)  demonstrating the power of this technique for detecting large-scale structure and its sub-components.
We used the DBSCAN function of the open source package fpc (``fixed point clusters'') implemented in the Comprehensive R Archive Network (CRAN). 
 The DBSCAN function in this package follows the procedure introduced by \citet{eks96}.

The DBSCAN technique
requires two user-defined input parameters: the minimum number of points (MinPts) which defines the minimum accepted number of the structure members, and the 
neighbourhood radius to search for the members associated with each
structure \citep{eks96}.
All of the data points have a classification as core, border or
noise. This method is mainly based on calculating the density of individual data points in the sample. The number of all data points within the radius ``Eps''  
is referred to as the density of that particular point. 
In the calculation of the density, the data point, itself, should be taken into account \citep{eks96}. In the case that the
number density of a point is greater than MinPts, that point is considered as a
core point.
A border point is not a core point but it is still within the Eps radius vicinity of a core
point and has lower number density value. The points which are not in the Eps radius proximity of the core points are classified as noise. 
All the steps in the DBSCAN procedure are described below: 
\begin{enumerate}
\item An arbitrary data point is chosen as a starting point. If the density of
that point exceeds the MinPts value, then that point is considered as a core
point, otherwise
another non-classified data point will be retrieved and the same procedure
is repeated. Once all the core points are found, the process finds the border and noise points until all the core and border points are classified.
Then, all the noise points will be removed from the sample and will not be further 
involved in processing. 
\item An arbitrarily core point is chosen and is allocated to the first structure. 
Then, any core point in the Eps-neighbourhood of that core point is also allocated to the first structure. 
This process recursively searches for other non-allocated core points within
the Eps radius around each core point and allocates them all to the first structure. 
This procedure is repeated till all the border points in the Eps-neighbourhood of
allocated core points are also allocated to the first structure.
The process is halted
when there are no core or border point left to be assigned to the structure.
\item A new core point which is not a member of any pre-detected cluster is retrieved and
step 2 is repeated and further clusters are discovered.
\item All the steps are repeated until all the data points are assigned to a cluster.
\end{enumerate}

%%%%%%%%%%%%%%%%%%%%%%%%%%%
\begin{figure}
\centering
%\hspace{-0.3cm}
 \includegraphics[scale=0.55,clip,trim=0mm 0mm 0mm 0mm]{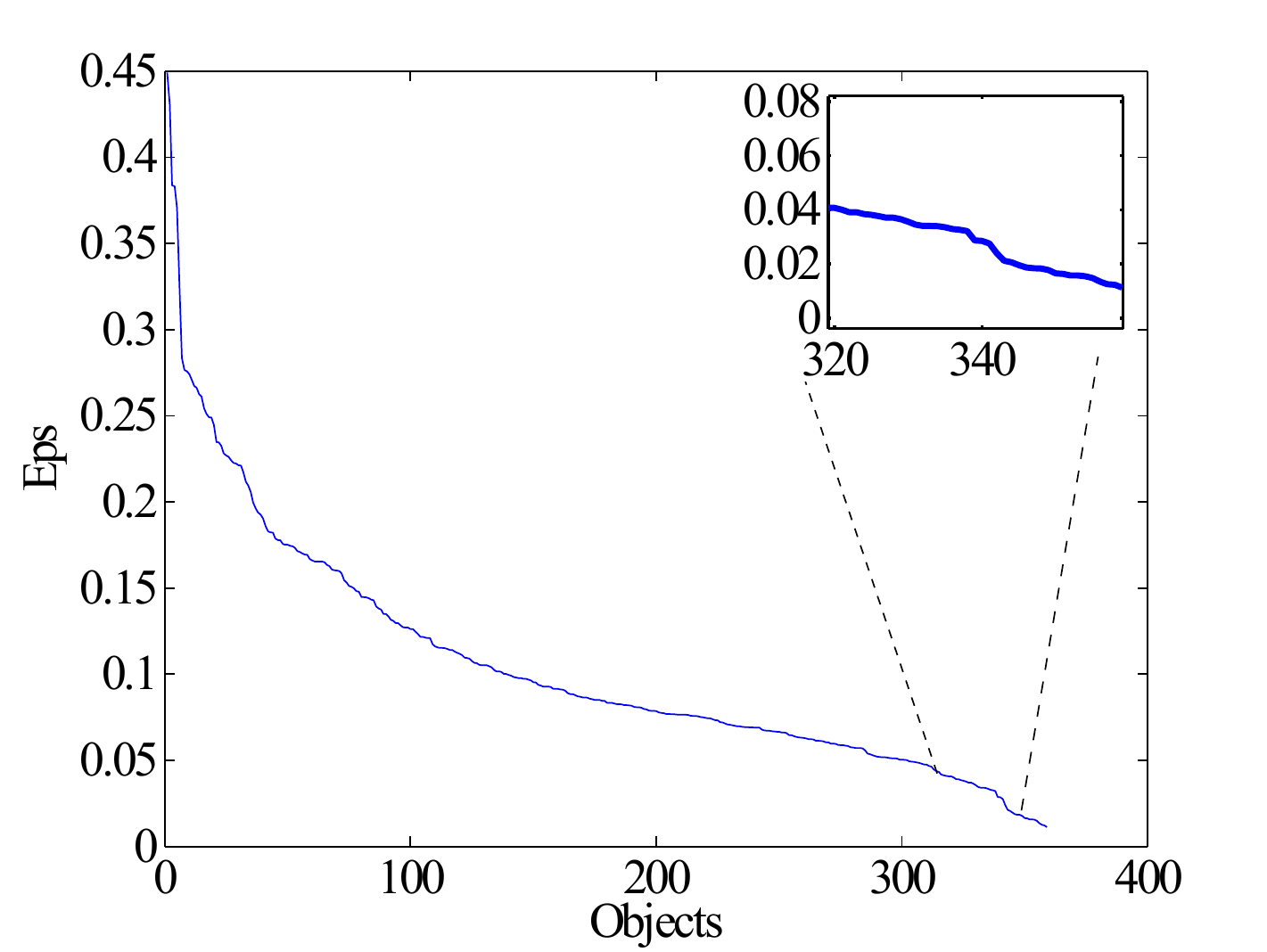}
 \caption[ The sorted k-dist plot with k=6]{ \small {The sorted k-dist plot with k=6. The zoomed inset
shows the change in the slope of this k-dist graph. The values which are
corresponding to strong changes in the slope are used as the Eps values for the DBSCAN
algorithm. The vertical axis has units of degrees and the horizontal axis has no physical
unit. The 6-dist sorted plot
shows an abrupt change around Eps= 0.027 ($\sim$ 270 kpc at redshift of cluster) and accordingly, the detected structures had proper scales.}}
\label{fig:kdist}
\end{figure}
%%%%%%%%%%%%%%%%%%%%%%%%%%%

The DBSCAN method has some advantages over the widely used Gaussian mixture model clustering algorithm in which
velocity groupings are identified by determining a combination of various fitted Gaussians to the dataset \citep{1994AJ....108.2348A}.
The Gaussian mixture clustering approach requires an initial guess of the total number of
groupings 
and the mean of each fitted Gaussian distribution.
However, DBSCAN does not need a prior guess of the
number of structures. In addition, DBSCAN is not restricted to detect any particular symmetric shape for the
groupings. However, the initial estimate of input parameters (MinPts, Eps) can broadly
affect the detected structures by DBSCAN. To avoid blind guessing of the input parameters, \citet{eks96} suggested to use a so called k-dist plot 
to choose the Eps value.

The k-dist plot can be constructed following two steps: 1) the distance to the k$^{th}$ (k = MinPts-1) nearest neighbour for each data point is calculated,
2) the distance values should be sorted in descending manner and plotted. Since the k value is MinPts-1, choosing a k $\le$ 6 leads to detection of many false groups with a small number of members.  
The best Eps value is the one
where the k-dist plot shows an abrupt change in the slope. The physical meaning is such that 
all the objects on the right side of the chosen Eps value are considered as core points. 
Thus, choosing a very small Eps value
leads to less detected structures whereas large Eps values 
tend to merge separate objects.
We have examined all the possible Eps values with fixed MinPts=7 in the DBSCAN 
application and found that Eps= 0.027 ($\sim$ 270 kpc at redshift of A3888) is the value of interest and corresponds to structures with physical scales e.g. galaxy groups. 
Figure~\ref{fig:kdist} shows the sorted
6-dist plot.

\subsection{Caustic Technique }
\label{subsec:subsec2}
In order to use the information of peculiar velocity of the galaxies in our cluster member selection procedure, the caustic
technique was used alongside the DBSCAN application. For this purpose, we used the CAUSTIC APP, an open source application initially developed by Ana Laura Serra 
\citep{CausticApp}. 

\citet{2013ApJ...768..116S} pointed out that ``the caustic technique \citep{1997ApJ...481..633D, 1999MNRAS.309..610D, 2009arXiv0901.0868D, 2011MNRAS.412..800S} 
identifies the escape velocity profile of galaxy 
clusters from their centre to radii as large as $3 R_{200}$''\footnote{$R_{200}$ is defined as the radius of a sphere 
where the average density is 200 times the current critical density 
of the Universe.}.
This radius is considered as the point in the cluster outskirts right before the warm-hot intergalactic
medium (WHIM) dominates the region \citep{2013SSRv..177..195R}. It is believed that the WHIM is the remaining
hot gas from galaxy formation activities in the past \citep{2013SSRv..177..195R}. 

The caustic method has been extensively employed 
to estimate the mass profile
of galaxy clusters at distances much larger than the virial radius ($\approx
R_{200}$) \citep{1989AJ.....98..755R, 1999MNRAS.309..610D, 2006AJ....132.1275R, 2011MNRAS.412..800S}.
However, this method is also a valuable technique to trace the imprints of the gravitational potential well of
the galaxy cluster at radii up to few megaparsecs. 

Galaxy clusters lie at the knots of filaments of galaxies in the cosmic web. 
Since gravity is a long-range acting force, galaxies at much
larger distances of the order of 10 Mpc will still be affected by the gravity of the cluster \citep{1997ApJ...481..633D}.
During mass accretion in the outskirts of the clusters, galaxies are accelerated by the cluster gravitational potential and consequently 
fall into the cluster. 
The aforementioned galaxies 
ultimately become a gravitationally bound member of the parent cluster. This process creates a characteristic trumpet-shaped
escape velocity profile/putative caustic
profile for all the galaxies which are interacting with the gravitational potential well of the cluster \citep{1987MNRAS.227....1K, 1989AJ.....98..755R,
1993ApJ...418..544V}.

The principal assumption in the caustic method is that clusters are spherical and symmetric systems, 
however, the cluster's morphology does not affect the caustic profile. 
The amplitude of the caustics, $A(r)$, is a
combination of escape velocity profile and the velocity anisotropy parameter,
$\beta (r)$, which is given by:
 \begin{equation}
 \beta (r) = 1-(\langle v_{\theta}^{2} \rangle + \langle v_{\phi}^{2}
\rangle)/2 \langle v_{r}^{2} \rangle\;
 \end{equation}
 where $ v_{\theta}, v_{\phi}$ and $v_{r} $
 are longitudinal, azimuthal and radial component of the galaxy velocity in the
volume $dr^{3}$ at position r \citep{1997ApJ...481..633D}.
The average squared velocity of galaxies in a sphere with radius r is: 
 \begin{equation}
 \langle v^{2} \rangle= \langle v_{LOS}^{2} \rangle g(\beta)\; .
 \end{equation}
 where $v_{LOS}$ is the line of sight component
 of the galaxy velocity and
 \begin{equation}
 g(\beta)= \frac{3-2\beta(r)}{1-\beta(r)}\; .
 \end{equation}
 The escape velocity at the radius r is $\langle v_{esc}^{2}(r) \rangle$ = -2$\phi(r)$
where $\phi(r)$ is the gravitational potential. If we assume that the caustic
amplitude, $A(r)$,
 is the LOS component of the escape velocity $A^{2}(r) = \langle v_{esc,LOS}^{2}
\rangle$ then we have:
 \begin{equation}
  -2\phi(r) = A^{2}(r)g(\beta) \equiv \phi_{B}(r)g(\beta)\; .
 \end{equation}
 This equation shows that the caustic amplitude, $A(r)$, is related to the potential well of the cluster .
 
To estimate $A(r)$, three major steps should be followed \citep{2011MNRAS.412..800S}:
 \begin{enumerate}
  \item Construction of a hierarchical 
  tree based on the calculation of projected binding energy for each pair of galaxies, presumed to have identical mass.
  \item Determining a threshold to terminate the growth of the hierarchical tree and find the candidate cluster members. 
  \item Calculating the cluster centre, galaxy number density, caustic profile and determining optimal cluster members.
  \end{enumerate}
At first, the projected pair-wise binding energy of all of the galaxies pairs are used to create a hierarchical tree.
In the second step, a threshold should be determined to cut the tree.

\begin{figure}
\centering
 \includegraphics[scale=0.58]{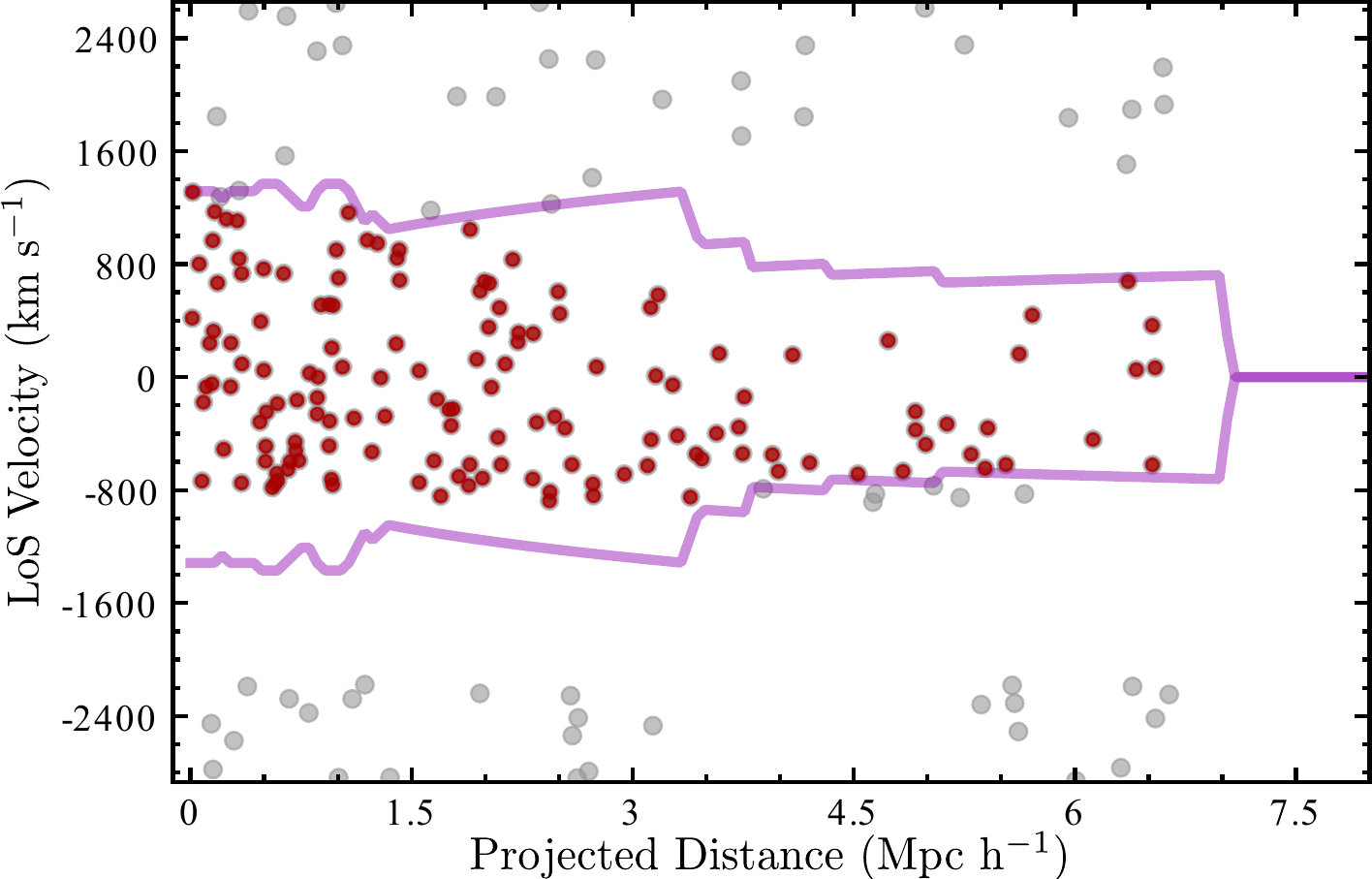}
 \caption[The caustic profile of A3888]{{ The caustic profile for A3888 (pink solid lines). The red (dark grey) dots show the cluster members defined by 
 the caustic technique. The grey dots are the interloper galaxies
 which are not gravitationally bound to A3888.}}
 \label{fig:caustic}
\end{figure}

\citet{1999MNRAS.309..610D} explained that in the hierarchical tree, there is a main branch that starts from the root and links ``nodes'' being places 
where galaxies are hanging.
Each group of galaxies hang from nodes that have a velocity dispersion $\sigma_{n}$. 
The distribution of velocity dispersions of the nodes shows a characteristic feature which is used to determine the cutting threshold.
The velocity dispersion $\sigma_{n}$ begins to drop
significantly due to less interloper contamination and then the velocity dispersion begins to flatten until it reaches a node associated 
with internal structures after which it shows a significant drop. 
\citet{1999MNRAS.309..610D} mentioned that most of the galaxies hang from node where the ``$\sigma$ plateau'' begins, are defined as candidate cluster members. 
In the last step, the cluster centre
and the galaxy density is calculated to estimate the LOS caustics. 
The caustic profile of A3888 is shown in Figure~\ref{fig:caustic}. Red (dark grey) dots represent the determined cluster members which reside within the caustics. The other galaxies in grey are not gravitationally bound to A3888.

%%%%%%%%%%%%%%%%%%%%%%%%%%%
%%%%%%%%%%%%%%%%%%%%%%%%%%%

The caustic technique
measures the escape velocity profile which can be used as a cluster member
allocation tool in the dynamical analysis of the clusters (e.g.
\citealt{2013ApJ...772..104O}, \citealt{2013ApJ...767...15R} and \citealt{2014ApJ...780..163O} ).
The caustic profiles of clusters carry invaluable information about each of the galaxies in the field of cluster.
The position of individual galaxies in the caustic profile reflects the time when infall towards the cluster started. Galaxies
which are passing the cluster core for the first time have
higher caustic amplitudes due to acceleration whereas
galaxies which have accreted at the same time that the core of the cluster was
forming, tend to be closer to the cluster centre 
and have lower caustic amplitudes \citep{2012ApJ...754...97H}.
The results of the DBSCAN and caustic technique applications are presented in the following sections.

\section{Structures in the Main Population of A3888}
\label{sec:sec7}
The colour-coded redshift histogram of the galaxies in a one degree radius around A3888, revealed that 
members of A3888 fall in the range of
0.12 $<$ z $<$ 0.0185. 
In Section \ref{sec:sec6}, we explained how the DBSCAN and caustic techniques were used to constrain 
the interlopers in the galaxy member selection process. These methods allowed
us to utilise both redshift and position information to determine the most
probable member candidates of A3888. 

Firstly, the DBSCAN
method was applied to the data in the redshift range 0.12 -- 0.0185 and consequently, 77 candidate galaxy members were identified. 
Further refinement with the caustic method
shrunk the number of cluster members to 67 galaxies. In order to verify that the
final results were not affected by the order of application, the reverse order (caustic-DBSCAN) was also examined. 
In the reverse order, the caustic technique was first applied to the data and
146 galaxies were identified as member candidates of the cluster A3888. Then, the DBSCAN method was applied  
to the caustic method results and 65 galaxies were identified as the final candidate
members of A3888. There were 61 galaxies in common in both orders of applications, four
extra galaxies were detected in the caustic-DBSCAN order and six further
galaxies in DBSCAN-caustic order. 
Because the caustic-DBSCAN order is
more conservative than the reverse order (DBSCAN-caustic) we combined both
results and consequently the total number of candidate galaxy members increased to 71. 

\begin{figure}
\centering
\includegraphics[scale=0.7]{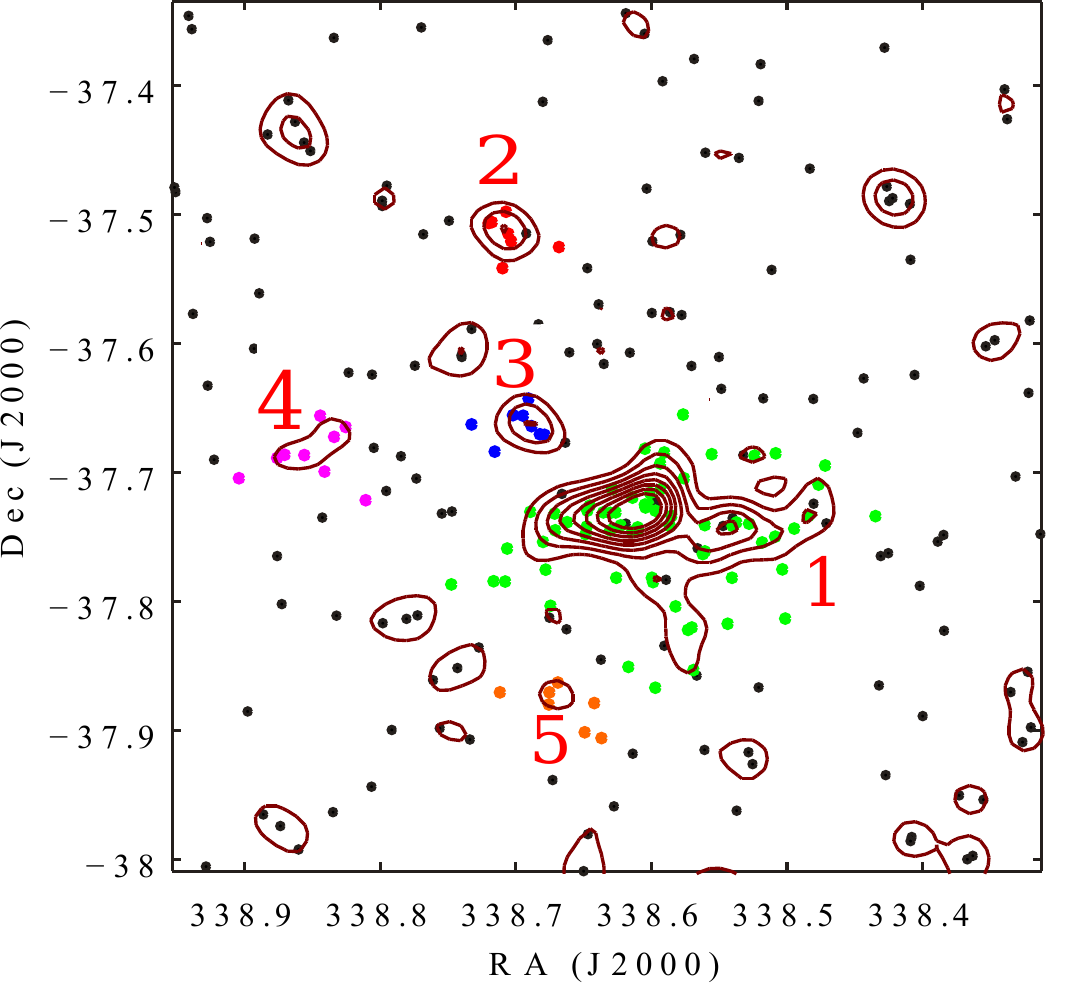}
\caption{ \small {The isoline plot of the groups and cluster A3888 detected by the DBSCAN method. 
The contours start at 10\% and continues in 10\% interval to 90\%. Dots labelled from 1--5 represent the main cluster A3888 and galaxy groups 2, 3, 4 and 5. See Table~\ref{tab:a3888groups} for more details.}}
\label{fig:dbscan-good}
\end{figure}

In addition, the DBSCAN method detected four other galaxy groupings
in the redshift range of A3888. 
The isoline plot of the cluster A3888 and the four detected galaxy grouping are shown in Figure~\ref{fig:dbscan-good}.
In the following discussion, details of A3888 and other detected structures in redshift slice 0.12 -- 0.0185 are given
and the five structures are also shown on Figure~\ref{fig:dbscan-good} as numbered groups of coloured points.
We have used a method presented by \citet{bfg90} to estimate the redshift and velocity dispersion of the detected structures. 
All the reported confidence intervals are calculated based on $\alpha$=0.05 (95$\%$ confidence level).
In Table~\ref{tab:a3888groups} details of the detected groups such as the rest frame velocity dispersion, and
observed redshift are given.

%%%%%%%%%%%%%%%%%%%%%%%%%%%
%%%%%%%%%%%%%%%%%%%%%%%%%%%

\begin{itemize}
\item Structure 1\\
This is the main cluster, A3888 with 71 members. The members are the combined resultant galaxies of caustic-DBSCAN and DBSCAN-caustic applications. 
The presence of an elongation along a East-West axis
might be indicative of sub-clustering in A3888. 
The estimated average redshift and rest frame velocity
dispersion of the cluster were calculated as 0.1535 $\pm$ 0.0009 and 1181 $\pm$ 197 km s$^{-1}$, respectively. Details of further substructure tests for the main cluster are given in Section \ref{sec:lee-a3888}.
\item Structure 2\\
This is a galaxy over-density located in the far North of A3888 and has 7 candidate members at a mean redshift 0.1406. 
This group was previously reported as a cluster candidate ($\epsilon$) by \citet{2001PhDT........74P} based on the redshift measurements of 5
galaxies. However, due to the small number of members, low rest frame velocity dispersion of 338 km s$^{-1}$, and lack of an obvious cD galaxy we classify this as a group. The caustic method showed that the galaxy members of this group
are not located between the caustics of A3888. Thus, this group is not gravitationally bound to the cluster.
\item Structure 3\\
This is a galaxy group in the North of cluster A3888 with 5 candidate members. 
The estimated mean redshift and rest frame velocity dispersion  are 0.1411 and 164 km s$^{-1}$, respectively. Using the caustic method revealed that the galaxies in this
group are not gravitationally
bound to the main cluster A3888. 
\item Structure 4\\
This structure with 9 members has two distinct velocity distributions and consequently, the rest frame velocity dispersion at 
mean redshift of 0.1529 was estimated as 1025 km s$^{-1}$ which is considerably higher than the expected velocity dispersion of galaxy groups (400--500 km s$^{-1}$). Additionally a visual inspection of the DSS red and blue images failed to suggest a real mass concentration.
As a result we believe this is not a real structure, but rather a false grouping.
\item Structure 5\\
This is a false positive structure featuring a sporadic velocity distribution. There is no evidence of Gaussianity in the velocity distribution. This suggests that
this detection is entirely contaminated by projection effects of galaxies which are distributed in a large redshift slice. 
The estimated mean redshift and rest frame velocity dispersion
were 0.1603 and 3232 km s$^{-1}$, respectively and again a visual inspection of the DSS images found no evidence of a real group.
\end{itemize}

%%%%%%%%%%%%%%%%%%%%%%%%%%%
\begin{table*}
 \centering
 \caption{Details of the detected galaxy groups and the cluster A3888 in the main population. 
All of the detected structures are in the redshift range of 0.12 -- 0.185. The upper and lower limits of the redshift and rest frame velocity dispersion represent the 
estimated $\alpha$=0.05, 95$\%$ confidence intervals.}
\begin{tabular}{cccccl}
\hline
\\[-10pt]
Structure ID&N$_{gal}$& z&$\sigma_{v,rest} [\mbox{km}$ s$^{-1}$] &Coord&Comment\\
\\[-10pt]
\hline
\\[-10pt]
1&71& 0.1535$^{+0.0009}_{-0.0009}$& 1181$^{+197}_{-197}$&22 34 31.01\hspace{2mm}-37 44 06& \small Cluster A3888\\
2&7& 0.1406$^{+0.0011}_{-0.0011}$&338$^{+386}_{-126}$  &22 34 52.01\hspace{2mm}-37 30 21& \small Galaxy group\\
3&5& 0.1411$^{+0.0008}_{-0.0008}$&164$^{+305}_{-66}$   &22 34 48.52\hspace{2mm}-37 39 20 &\small Galaxy group\\
4&9&0.1529$^{+0.0037}_{-0.0037}$&1025$^{+402}_{-402}$& N/A&\small False galaxy group \\
5&7&0.1603$^{+0.0108}_{-0.0111}$&3232$^{+301}_{-1978}$& N/A&\small False galaxy group \\
\\[-10pt]
\hline
\end{tabular}
\label{tab:a3888groups}
\end{table*}
%%%%%%%%%%%%%%%%%%%%%%%%%

%%%%%%%%%%%%%%%%%%%%%%%%%%%
\begin{figure}
 \centering
\hspace{-5mm}
\includegraphics[scale=0.45]{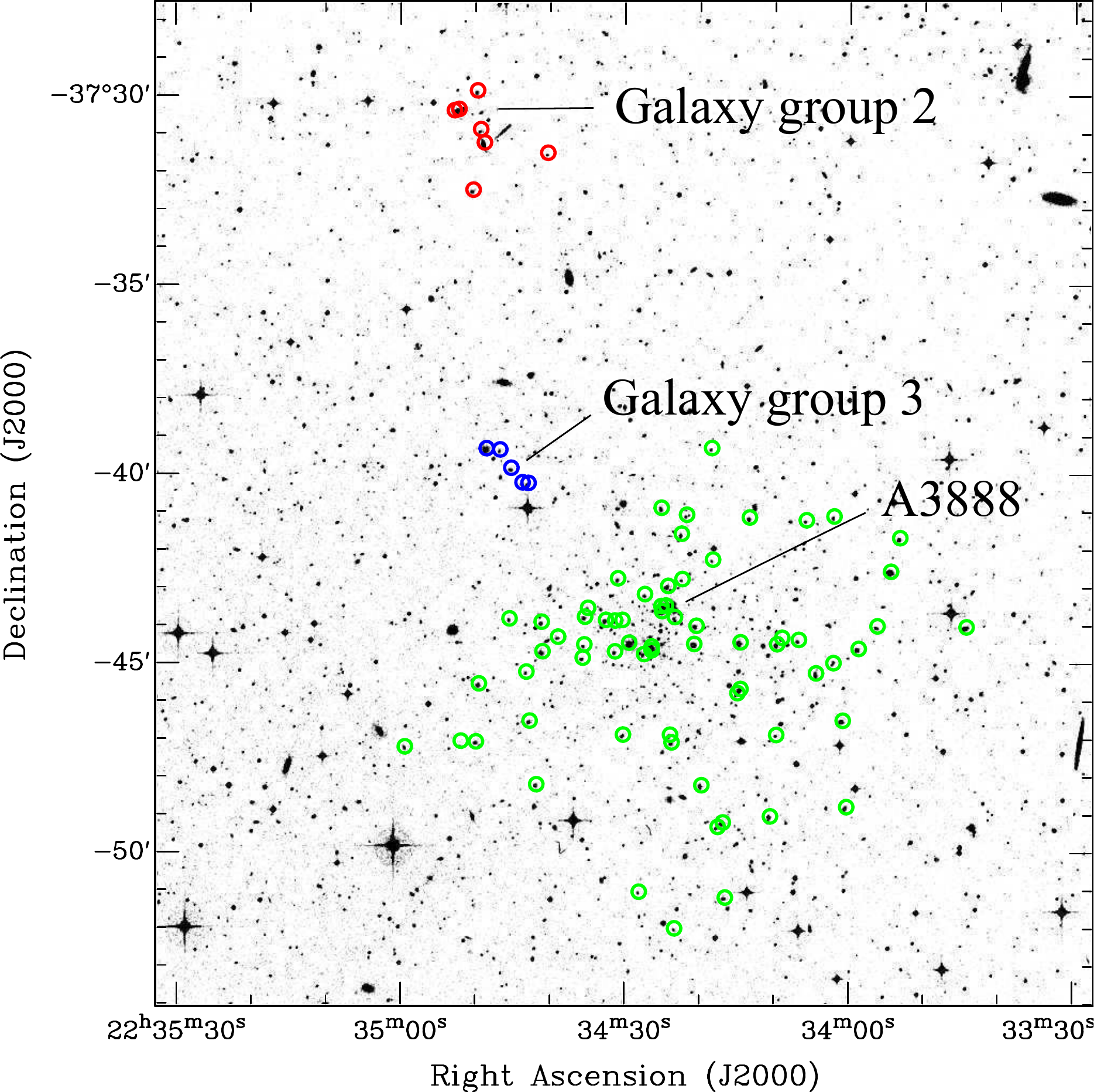}
\caption {Distribution of member galaxies of cluster A3888 and the two detected groups (Structures 2 $\&$ 3). A3888 and galaxy groups 2 and 3 are demonstrated 
with green (light grey), red (medium grey) and blue (black) open circles, respectively, on the DSS blue image.}
\label{fig:A3888-dss}
\end{figure}
%%%%%%%%%%%%%%%%%%%%%%%%%%%

%%%%%%%%%%%%%%%%%%%%%%%%%%%
\begin{figure}
\centering
 \includegraphics[scale=0.6]{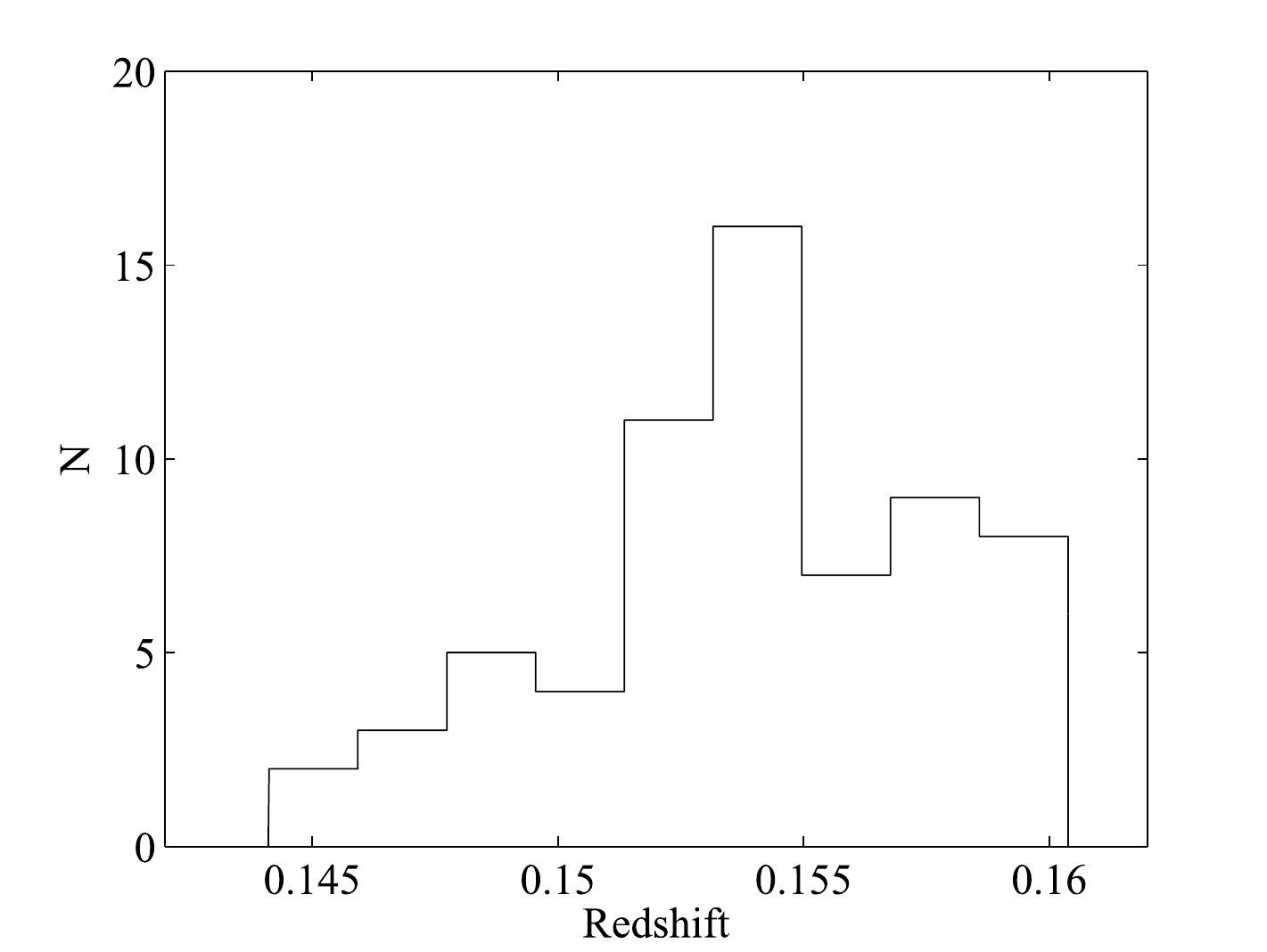}
 \caption{Redshift histogram of the 71 possible members of A3888 shown with a bin width of 540 km s$^{-1}$.
 A departure from Gaussianity is evident around z$\sim$ 0.157.}
\label{fig:71memberhist}
\end{figure}
%%%%%%%%%%%%%%%%%%%%%%%%%%%

In Figure~\ref{fig:A3888-dss} the spatial distribution of the member galaxies of 
A3888 and the detected galaxy groups (structures 2 and 3) are shown. The background image is the DSS blue image.

\section{Substructure Test of A3888}
\label{sec:lee-a3888}
\begin{figure*}
 \centering
 \includegraphics[scale=1.3,clip,trim=5mm 5mm 7mm 5mm]{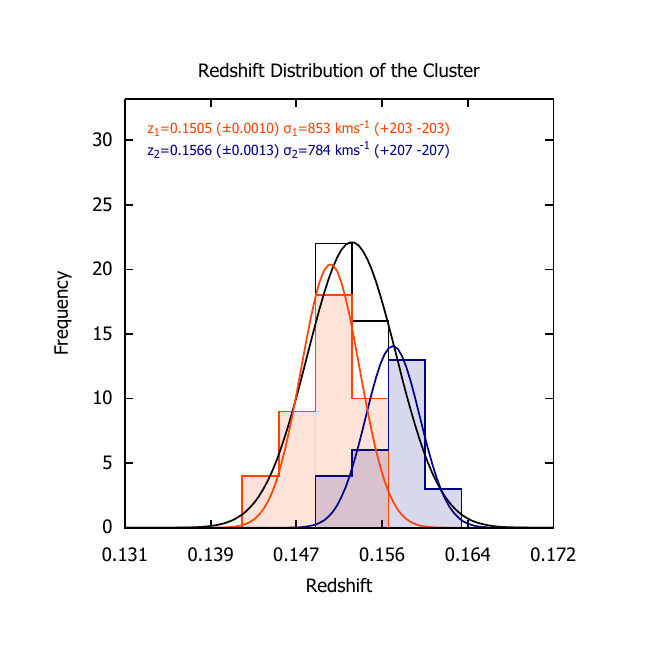}
 \includegraphics[scale=1.3,clip,trim=0mm 5mm 5mm 5mm]{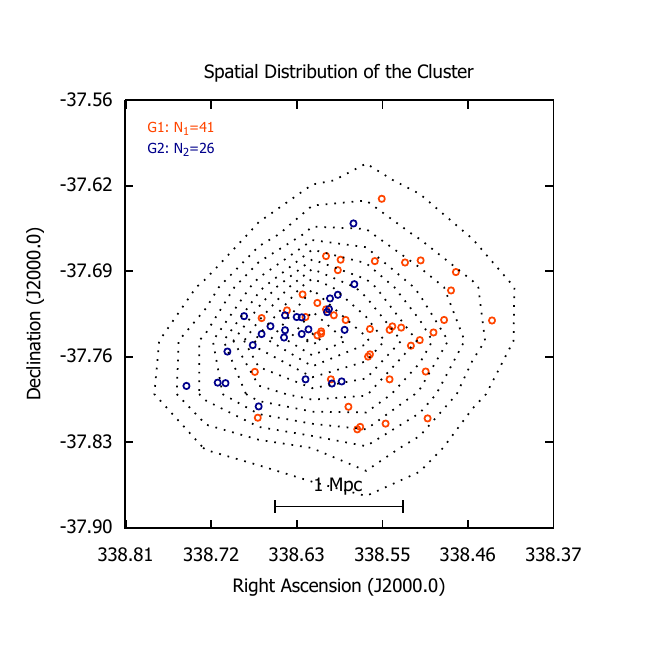}
 \vspace{-3mm}
 \caption{Results of the Lee-Fitchett 3D substructure test of A3888. 
 The cluster is bimodal and the two subgroups are denoted as G1 in red (light grey) and G2 in blue (dark grey).
 Left plot: histogram of all the members, G1, and G2 are shown with black, 
 red (light grey), and blue (dark grey) lines, respectively. The estimated redshift and rest frame velocity dispersion for G1 and G2 are given in red and blue (light and dark grey), respectively. 
 Right plot: Spatial distribution of the detected groups G1 and G2. The number of candidate members of G1 and G2 are shown in red and blue (light and dark grey), respectively. }
 \label{fig:lee}
\end{figure*}

In Figure~\ref{fig:71memberhist} the redshift histogram of the 71 final
cluster member candidates is plotted. A departure from Gaussianity is evident around redshift 0.157. In order to verify that this peak in the histogram is statistically significant,
 we performed a Lee-Fitchett \citep{1979ASA.74..708, 1988MNRAS.230..161F} 3D substructure test using the CALYPSO package (Dehghan \& Johnston-Hollitt in prep.). 
 A clipping procedure in the software has automatically rejected four of our 71 candidate member galaxies. According to the substructure test result, A3888 is bimodal.  
 The substructure significance was 99.8$\pm{0.2}\%$ (99.00\% confidence level)
which is equivalent to a 3.16-sigma detection.
However, due to the relatively low redshift sampling of the region, this result is likely subject to change as more redshifts become available in the future.
 Figure ~\ref{fig:lee} shows the results of the Lee-Fitchett 3D test for A3888.
 The detected subgroups named G1 and G2 have average redshifts of 0.1505 and 0.1566, respectively. The rest frame velocity dispersion of G1 and G2 were
 measured as 853$\pm{203}$ and 784$\pm{207}$ km s$^{-1}$ at 95\% confidence level. Notably the cluster is split into the two subgroups along the same East-West axis as the overall density elongation.
 Table~\ref{tab:lee-sub} gives the properties of the two subgroups detected here.

%%%%%%%%%%%%%%%%%%%%%%%%%%%
%%%%%%%%%%%%%%%%%%%%%%%%%%%

%%%%%%%%%%%%%%%%%%%%%%%%%%%
\begin{table}
\centering
\caption{Details of the detected substructures in A3888.} 
\label{tab:lee-sub}
 \begin{tabular}{cccc}
 \hline
 \\[-10pt]
 Groups   &   N$_{gal}$ &  z    &   $\sigma_{v,rest} [\mbox{km}$ s$^{-1}$] \\
 \\[-10pt]
 \hline
 \\[-10pt]
G1 &41  & 0.1505$\pm{0.0010}$ & 853$\pm{203}$ \\            
G2 &26 &  0.1566$\pm{0.0013}$ & 784$\pm{207}$\\
\\[-10pt]
\hline
 \end{tabular}
\end{table}
%%%%%%%%%%%%%%%%%%%%%%%%%%%

We stress that the cumulative redshift
coverage is only up to 30\% within 1.5 Mpc radius around the cluster  
where the 71 candidate members reside. Thus,
this substructure test is subject to change with improved spectroscopic redshift sampling of the region.

\section{Morphological Content of A3888}
\label{sec:sec9}

We have used the spectral atlas of galaxies presented by \citet{1996ApJ...467...38K} to determine the
morphology of the galaxies in our AAOmega sample throughout the redshift determination by RUNZ. All sets of ASCII templates cover the
ultraviolet to
near infra-red spectral range
with the wavelength range of 1235--9935{\AA}. These spectral templates are
ideal to identify various galaxy morphologies from elliptical to late
type spirals. The full details of the templates for different galaxy
morphology are found in \citet{1996ApJ...467...38K}.
All of the spectra were visually inspected to confirm the galaxy morphologies 
determined by RUNZ. We have also used the spectrophotometric atlas of galaxies presented by \citet{1992ApJS...79..255K}
to compare our spectra with the reference spectra provided in this atlas. 
We found that the majority of our observed galaxies across the field of AAOmega were late-type spirals (S$_{b}$ and
S$_{c}$).
However, these spiral galaxies were mainly field galaxies scattered
in various redshift slices. As it is expected, most of the early type galaxies were found in the 
over-dense regions particularly within a 1 Mpc radius around the core of A3888.

Understanding the evolution of early/late type galaxies in clusters has been one of the major interests in optical astronomy.
Recent large spectroscopic surveys (see Section \ref{sec:sec1}), have provided an ideal opportunity to 
explore the morphological content of galaxy clusters and probe their evolution. For instance, 
\citet{2003PASJ...55..739G} studied 514 galaxy clusters in the redshift range 0.02 $<$ z
$<$ 0.3 from the SDSS. According to their findings, the fraction of blue 
galaxies at the redshift of A3888 (z = 0.1535) should be around 15$\%$. 
Later on, \citet{2009ApJ...697L.137P} investigated the correlation between 
morphological content of clusters with the cluster's global properties such as X-ray luminosity and velocity dispersion.
They studied the
Wide-Field Nearby Galaxy-cluster Survey clusters (WINGS) extensively to examine the effect of cluster
properties and environment on the evolution of different galaxy morphology in clusters.
WINGS is a multi wavelength photometric and spectroscopic survey of
77 galaxy clusters at 0.04 $<$ z $<$ 0.07 \citep{2006A&A...445..805F}. 
They additionally used 15 high-redshift clusters in former publications to accurately quantify any observed
correlation. \citet{2009ApJ...697L.137P} found that in general, at lower redshifts the
morphological content of galaxy
clusters are 23:44:33 corresponding to spiral, S0 and elliptical galaxies respectively.
Moreover, they found that the spiral fraction of galaxy clusters, inversely correlates with the X-ray luminosity of the parent cluster.

To expand our optical analysis of A3888 and to shed light on the discrepancy associated with the X-ray luminosity, we have investigated the morphological content of A3888 and 
also verified the location of A3888 with respect to current proposed correlations between the morphological segregation in clusters 
and the global properties of the parent cluster.
Since the majority of redshifts of candidate members of A3888
come from the spectroscopic observations in the LARCS project \citep{2006MNRAS.366..645P},
we used the B-R colour reported in the LARCS spectroscopic catalogue to derive the spectral typing of the galaxies in our final redshift sample.
 This treatment has allowed us to supplement our AAOmega spectral typing information to determine the morphological fraction of the cluster A3888. 
 \citet{2006MNRAS.366..645P} reported that the colour of the emission and absorption line galaxies are (B-R)$<$ 1.6 
and (B-R)$>$ 2 respectively in their sample, thus galaxies with B-R
colour between 1.6 and 2 could be either an absorption or an emission line.
 
\begin{figure}
\centering
\hspace{-5mm}
\includegraphics[scale=0.8]{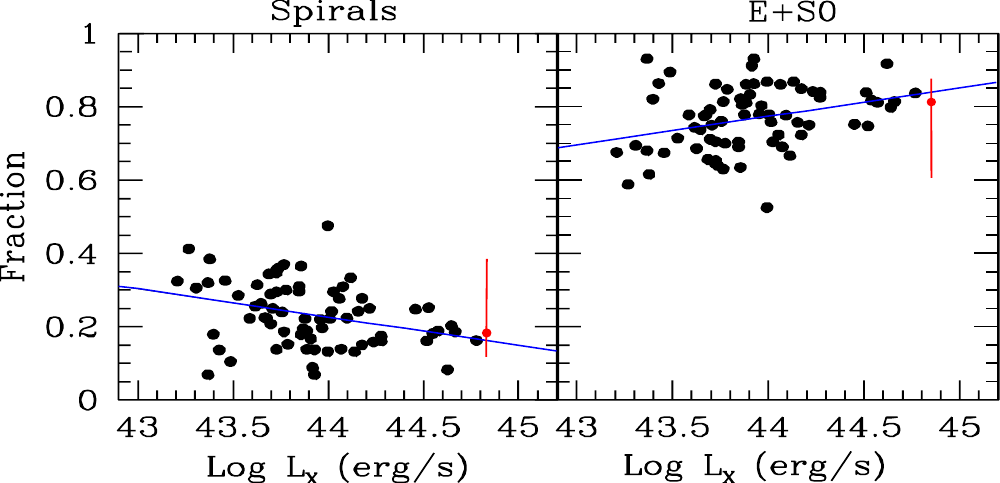}\\
\vspace{0.5cm}
\includegraphics[scale=.855]{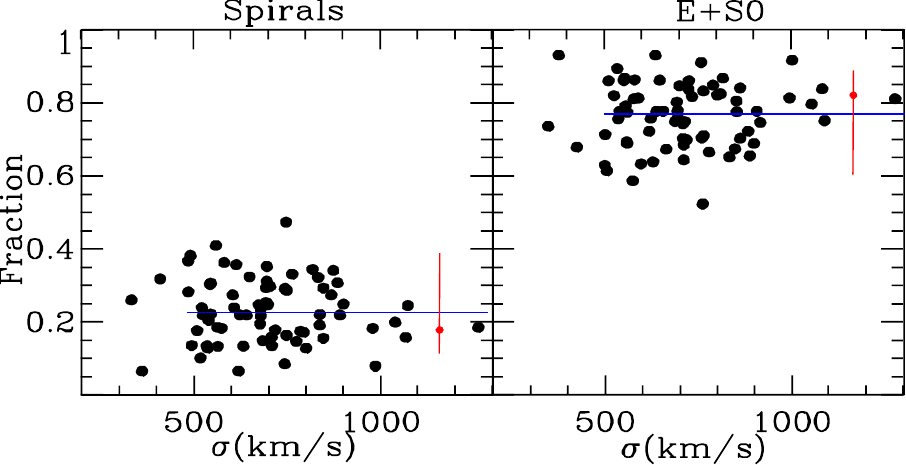}
\caption{\small {Top panels: the morphological fraction correlation with the X-ray luminosity
of the parent cluster is shown with a blue line \citep{2009ApJ...697L.137P}. Bottom panels: 
the morphological fraction correlation with the cluster's rest frame 
velocity dispersion is shown with a blue line \citep{2009ApJ...697L.137P}. In all the plots, 
the location of A3888 is annotated with the filled red dot and a vertical line to denote the uncertainty. The images are
adapted from \citet{2009ApJ...697L.137P} with permission.}}
\label{fig:morphlx}
\end{figure}

In total, the 71 candidate
members of the cluster were comprised of 9 emission line galaxies, 43 absorption line
galaxies, 7 galaxies with 1.6 $<$B-R$<$ 2 in
the LARCS catalogue and the remaining 12 members had
no morphological or spectral typing data available in the literature. Thus of the 52 galaxies with confirmed morphological types, $\sim$ 17\% (9/52) are emission line galaxies and $\sim$
83\% (43/52) are absorption line objects.
Additionally, for the 19 remaining galaxies (12 without morphological information and 7 with 1.6 $<$B-R$<$ 2) one could assume these could be 100\% emission or
absorption line to derive the upper and lower limits of morphological content of the cluster.
According to these assumptions the estimated emission and absorption fractions are 0.17$_{0.13}^{0.40}$ and 0.83$_{0.60}^{0.88}$, respectively. In Figure~\ref{fig:morphlx}, the two top plots show the correlation between the X-ray luminosity of the parent cluster and 
the morphological content of the cluster as the fraction of galaxy morphological type in \citep{2009ApJ...697L.137P}. The location of A3888 is marked with a red filled 
dot with errors from the limits of the morphological fraction described above. 
The bottom plots indicate the correlation between the rest frame velocity dispersion and the morphological fraction of the clusters.
As can be seen from the plot, the morphological content of A3888 is 
in agreement with the results derived by \citet{2009ApJ...697L.137P}.
Although our redshift coverage is not
similar to aforementioned galaxy cluster surveys, our results
suggest that the morphological fraction of galaxies in A3888
are in line with the predicted values in large surveys. 
Additionally we see that the position of A3888 obtained using the X-ray luminosity of \citet{2009A&A...498..361P} 
is consistent with known cluster properties and using either of the previously reported L$_x$ 
values moves the position of the cluster to a less consistent position. However,
we note that the uncertainties on the morphological content make this only a weak argument. 
Nevertheless it does lend further credence to the X-ray luminosity derived by \citet{2009A&A...498..361P}.

%%%%%%%%%%%%%%%%%%%%%%%%%%%
%%%%%%%%%%%%%%%%%%%%%%%%%%%

\section{Foreground and Background Structures}
\label{sec:sec8}
According to the colour-coded histogram (see Figure~\ref{fig:colorhist}), 
there are other fore/background populations of galaxies in a one degree radius around A3888. 
The DBSCAN method was applied to all of the separate 
populations distributed in the redshift histogram. According to the DBSCAN results, galaxy populations 
in the light-blue (0.053 $\le$ z $\le$ 0.0875) and red (0.185 $\le$ z $\le$ 0.222) regions in the colour-coded histogram 
(Figure~\ref{fig:colorhist}) host galaxy clusters and galaxy groups.

Examination of the DSS blue image revealed existence of three cD galaxies in the field with conspicuous
 mass concentrations around them. Interestingly, the application of the DBSCAN to galaxies in the redshift 
 range 0.053 $\le$ z $\le$ 0.0875 (the light blue grouping on Figure~\ref{fig:colorhist})
 confirmed the existence of three galaxy clusters around the aforementioned cD galaxies. Thus, the DBSCAN results 
were consistent with the visual inspection of the DSS blue image and hereafter these three clusters are denoted as Structures 6,7 and 8.
Examination of the redshifts between 0.185 and 0.222 (red population on Figure~\ref{fig:colorhist}) 
showed the presence of an additional galaxy group, denoted here as Structure 9.

Table~\ref{tab:sub2groups} gives the summary of the DBSCAN results on these over-densities and further narrows the ranges over which the structures occur.
We also list an IAU name for each new cluster detected.
On account of poor
redshift
coverage, determining all of the members of the detected clusters and galaxy groups is not feasible and all
the analysis is based on the available limited measured redshifts.

%%%%%%%%%%%%%%%%%%%%%%%%%%%%%%%%%%%%%%%%%%%%%%%%%%%%%%%%%%%%%%%%%%

\begin{figure}
\centering
\includegraphics[scale=0.43]{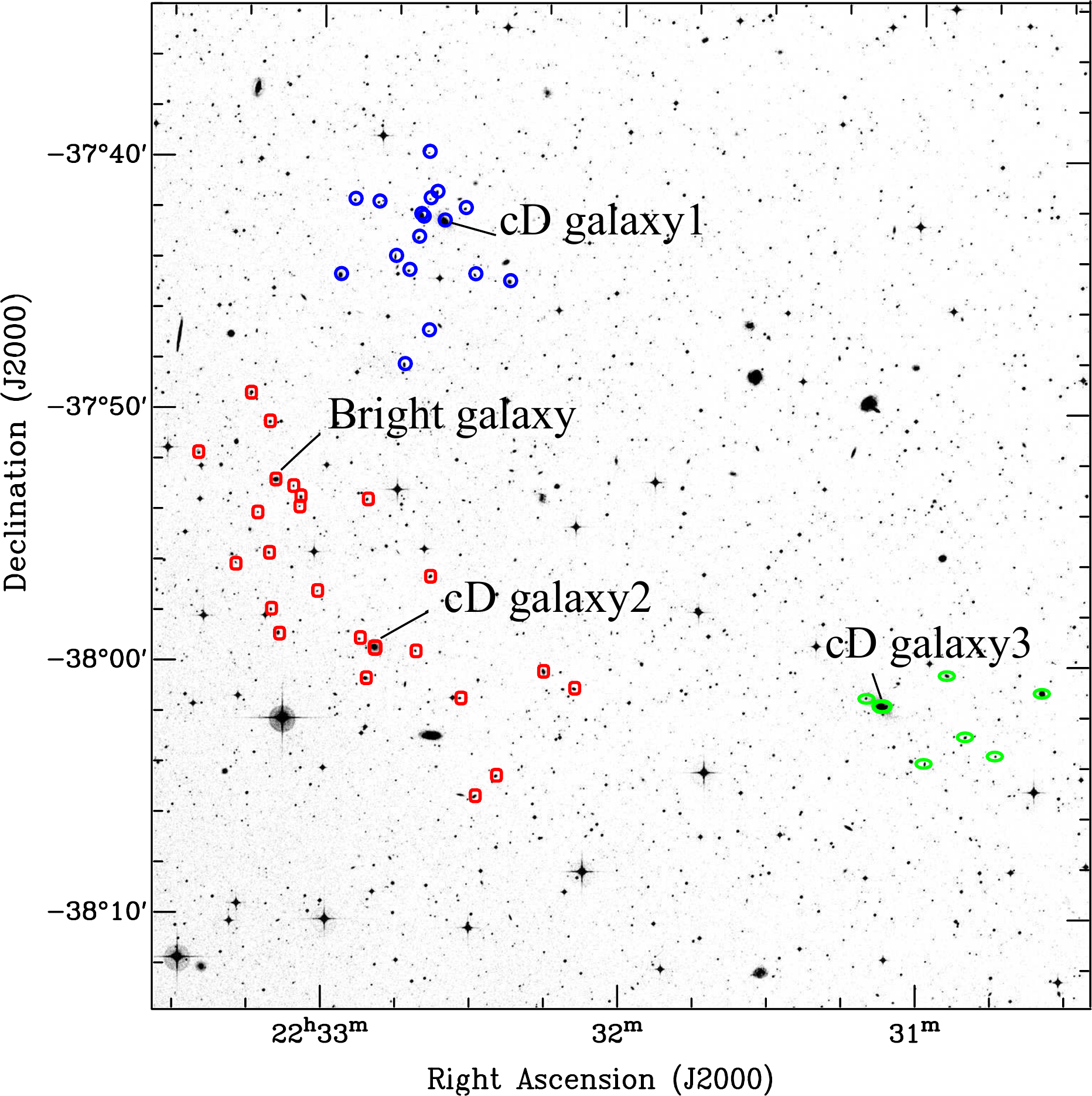}
\caption{The spatial distribution of galaxies in the detected
structures. The cD galaxies of detected galaxy clusters 
are annotated on the plot. Structure 6 (SJD J2232.4-3742) is shown with blue circles with labelled cD galaxy 1, 
Structure 7 (SJD J2232.5-3759) is given by red squares with labelled cD galaxy 2 and Structure 8 (SJD J2231.1-3801) 
is shown with green ellipses with labelled cD galaxy 3.
The DSS blue is the background image.}
\label{fig:sub2-all-groups}
\end{figure}

%%%%%%%%%%%%%%%%%%%%%%%%%%%%%%%%%%%%%%%

\begin{figure}
\centering
\includegraphics[scale=0.75, clip, trim=15mm 2mm 0 7mm]{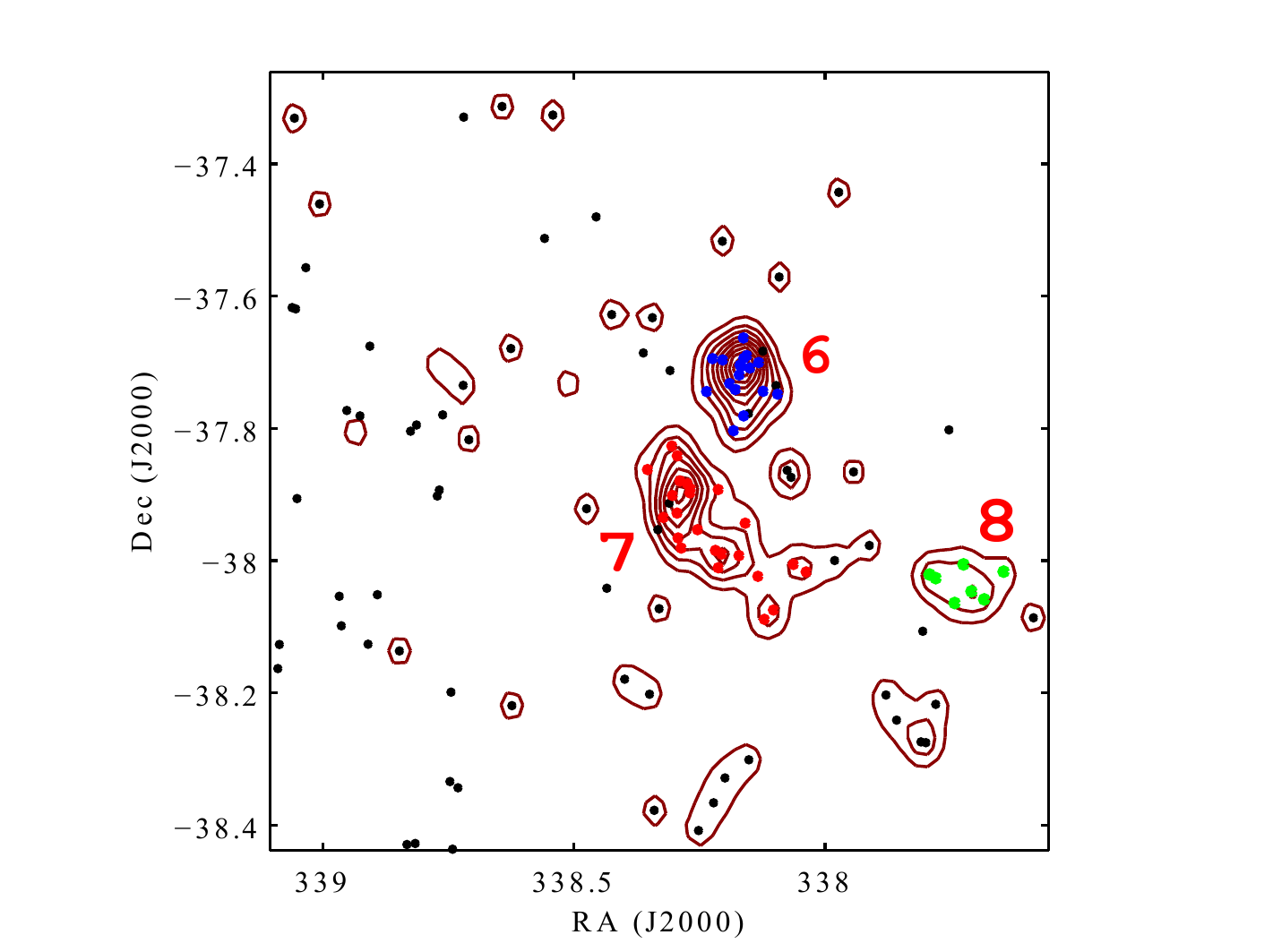}
\caption {The isoline plot of the galaxies in redshift slice of
0.0717--0.0856. The contour starts from 10\% and continues with 10\% interval to 90\%.}
\label{fig:sub2cl-dbscan-contour}
\end{figure}

%%%%%%%%%%%%%%%%%%%%%%%%%%%%%%%%%%%%%%%%%%%%%%%%%%%%%%%
\begin{figure*}
 \centering
 \hspace{-0.3cm}
\includegraphics[scale=0.51]{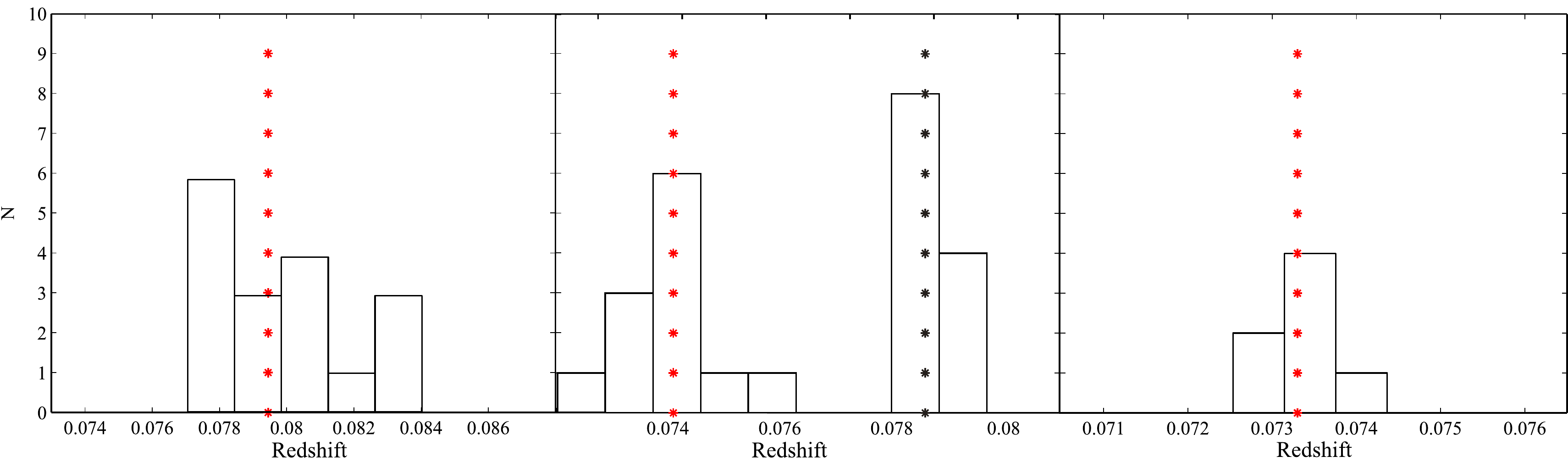}
 \caption{Redshift histogram of the clusters found in the redshift range 0.0717--0.0856. 
 The red (grey) vertical dotted lines represent the redshift of the cD galaxy in each
cluster. The black dotted line represents the redshift of the BCG.
 The redshift histograms of Structures 6 (SJD J2232.4-3742), 7 (SJD J2232.5-3759) and 8 (SJD J2231.1-3801) are shown from left to right, respectively.}
 \label{fig:sub2clusterhist}
\end{figure*}

\begin{figure}
 \centering
 \includegraphics[scale=0.44]{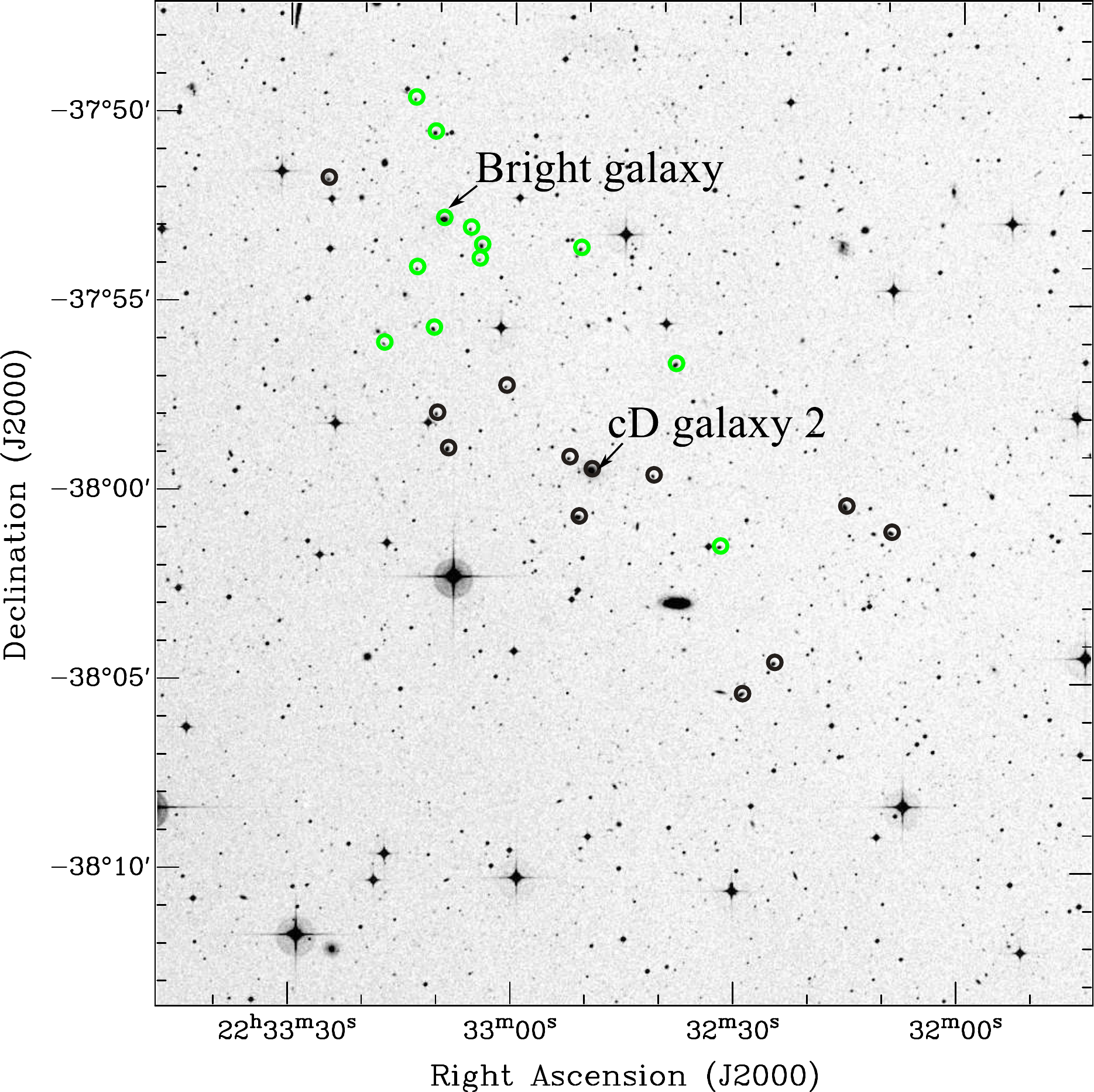}
 \caption{Redshift distribution of Structure 7 (SJD J2232.5-3759) based on the velocity gap in the
redshift histogram (see the middle plot in Figure~\ref{fig:sub2clusterhist}). The black circles represent the galaxy population of the left side of the 
 histogram in the redshift range of 0.073--0.077 and the light green circles demonstrate
the galaxies in the right side of the histogram in the redshift range of 0.077--0.080. The DSS blue is the background image.}
\label{fig:sub2g2-merging}
\end{figure}

\begin{figure}
\centering
\hspace{-5mm}
\includegraphics[scale=0.7, clip, trim= 0 0 0 2mm]{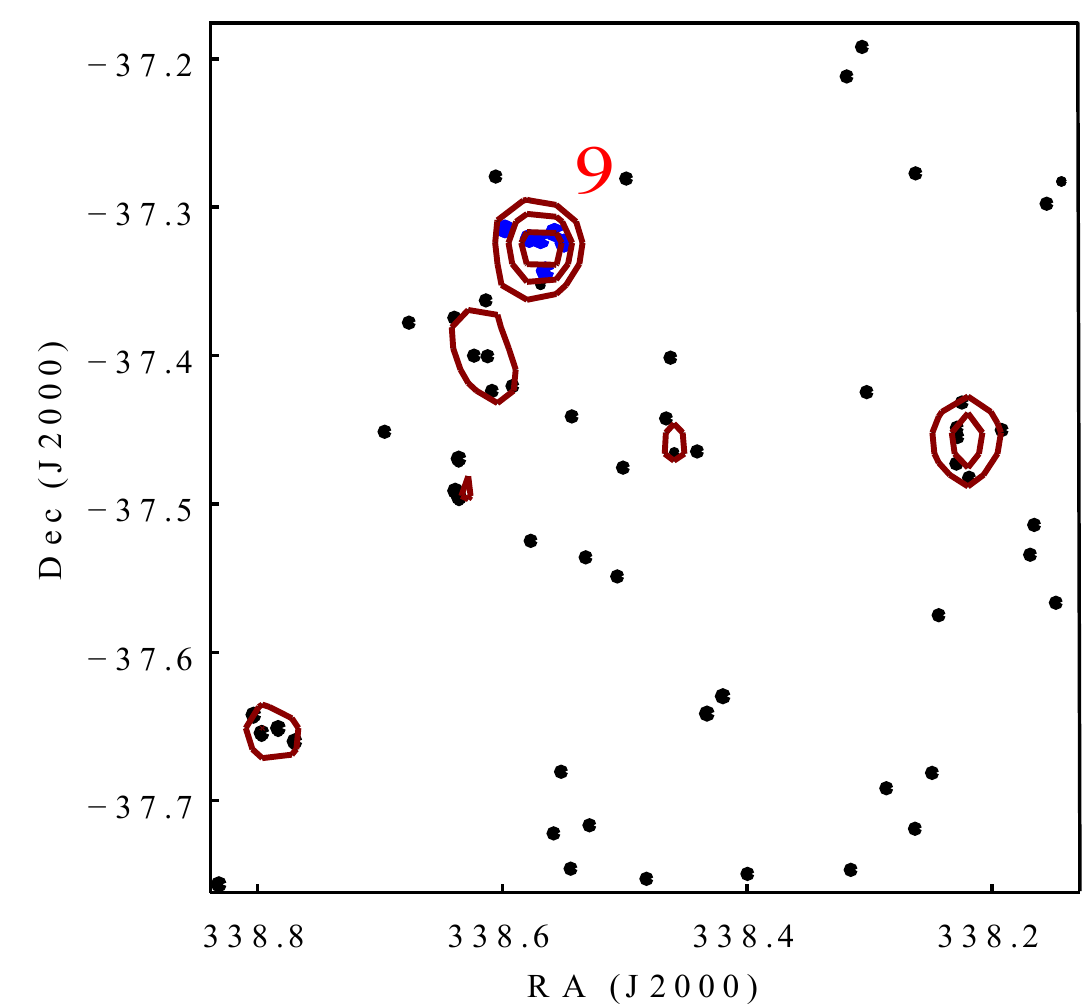}
\caption{ Contour plot of Structure 9 (SJD J2234.2-3719). The contour levels are 50\%, 70\% and 90\%.}
\label{fig:sub5coutour}
\end{figure}

%%%%%%%%%%%%%%%%%%%%%%%%%%%%%%%%%%%%%%%%%%%%
\begin{table*}
\centering
\caption{Details of fore/back ground structures detected by the DBSCAN method in the
field of A3888. The upper and lower limits of the redshift and velocity dispersion represent the 
estimated $\alpha$=0.05, 95$\%$ confidence intervals.}
\begin{threeparttable}
\begin{tabular}{ccccccc}
 \hline
 \\[-10pt]
 \small Structure ID &IAU Name&z$_{cD}$& N$_{gal}$ & z& $\sigma_{v,rest}$ [$\mbox{km}$ s$^{-1}$]&cD coord \\
 \\[-10pt]
 \hline
 \\[-10pt]
6& SJD J2232.4-3742&0.0795& 17& 0.0792$^{+0.0013}_{-0.0013}$& 730$^{+363}_{-363}$&22 32 35\hspace{1mm}-37 42 28.93\\
7& SJD J2232.5-3759&0.0741& 24& 0.0765$^{+0.0022}_{-0.0022}$& 727$^{+592}_{-592}$&22 32 49\hspace{1mm}-37 59 24.91\\
8&SJD J2231.1-3801&0.0733& 7&  0.0732$^{+0.0005}_{-0.0005}$ &128$_{-19}^{+209}$&22 31 07\hspace{1mm}-38 01 35.95\\
9&SJD J2234.2-3719&$\ldots$&6&0.2006$^{+0.0020}_{-0.0020}$&420$^{+541}_{-175}$&22 34 16\hspace{1mm}-37 19 25.01\tnote{a} \\
\\[-10pt]
\hline
 \end{tabular}
 \begin{tablenotes}
            \item[a] The coordinate of the brightest galaxy.
        \end{tablenotes}
     \end{threeparttable}
\label{tab:sub2groups}
\end{table*}
%%%%%%%%%%%%%%%%%%%%%%%%%%%%%%%%%%%%%%%%%
 As before, we used the method of \citet{bfg90} for the calculations of
velocity dispersions and redshifts of the detected structures. All of the confidence intervals  
were estimated based on 95$\%$ confidence level ($\alpha$=0.05).
\begin{itemize}
 \item Structure 6 -- SJD J2232.4-3742\\
  This structure is a galaxy cluster which has 17 candidate members. The cD galaxy has a 
  redshift of 0.0795 and is denoted as cD galaxy 1 on Figure~\ref{fig:sub2-all-groups}. 
  All of the candidate members of this cluster are annotated with 
blue open circles on the DSS blue image in Figure~\ref{fig:sub2-all-groups}. 
The estimated average redshift and rest frame velocity dispersion 
 of the cluster are 0.0792 and 730 km s$^{-1}$, respectively. The contour plot of this structure is shown in Figure~\ref{fig:sub2cl-dbscan-contour} (blue dots with label 6).
 In the left plot in Figure~\ref{fig:sub2clusterhist} the redshift histogram of this
cluster is shown. The details of this structure is given in Table~\ref{tab:sub2groups} and the member galaxies are repoted in Table~\ref{tab:new-clust}. 
We stress that the velocity dispersion might have been underestimated because of
poor sampling in this region. 
 \item Structure 7 -- SJD J2232.5-3759\\
 This structure is a galaxy cluster which has 24 candidate members with a cD galaxy with 
 redshift of 0.0741 which is denoted as cD galaxy 2 on Figure~\ref{fig:sub2-all-groups}. 
 The candidate members with red open squares are shown in Figure~\ref{fig:sub2-all-groups}.
The average redshift and the rest frame velocity
dispersion
 of this cluster is 0.0765 and 727 km s$^{-1}$, respectively. 
 The contour plot of this structure is shown in Figure~\ref{fig:sub2cl-dbscan-contour} (red dots with label 7).
 The middle plot in Figure~\ref{fig:sub2clusterhist} shows the redshift histogram of this
cluster with 
two well-separated redshift distribution which might explain the large estimated intervals of the velocity dispersion. The left peak 
 in the histogram is consistent with the redshift of the cD galaxy.
Additionally, the right peak in the histogram is in line with the redshift of a 
 bright galaxy (z = 0.0786) which is probably the Brightest Cluster Galaxy (BCG) of an interacting cluster and is also denoted as 
 ``Bright galaxy'' in Figure~\ref{fig:sub2-all-groups}. 
 The projected distance (estimated at the mean redshift of both galaxies) between these
two galaxies is 695 kpc. Figure~\ref{fig:sub2g2-merging} shows the
two populations of galaxies with different velocity distributions according to the histogram shown in middle plot in Figure~\ref{fig:sub2clusterhist}.
The redshift distribution might be a signpost of an dynamical interaction
between a galaxy cluster and a group or another cluster of galaxies. 
Table~\ref{tab:sub2groups} gives details of this structure and the member galaxies are repoted in Table~\ref{tab:new-clust}. Further analysis
requires greater redshift sampling of this region. 
 \item Structure 8 -- SJD J2231.1-3801\\
 This is a galaxy cluster with 7 members which has an average redshift of 0.0732 and hosts a cD galaxy with z= 0.0733 which 
 is denoted as cD galaxy 3 in Figure~\ref{fig:sub2-all-groups}. 
 The candidate members are annotated with green open ellipses in Figure~\ref{fig:sub2-all-groups}. 
 The contour plot of this structure is shown in Figure~\ref{fig:sub2cl-dbscan-contour} (green dots with label 8).
 The right plot in Figure~\ref{fig:sub2clusterhist} shows the redshift histogram of this
cluster. The redshift of the cD galaxy is in line with the peak of the
histogram. On account of insufficient redshift coverage the rest frame velocity
dispersion (128 km s$^{-1}$) will be highly underestimated, but is sufficient for a group.
Table~\ref{tab:sub2groups} gives details of this structure and the member galaxies are repoted in Table~\ref{tab:new-clust}. 
\item Structure 9 -- SJD J2234.2-3719\\
There is another galaxy group found in the redshift slice of 0.1867--0.2162. This
is a very small group of merely six galaxies. The average redshift and rest frame velocity dispersion of this group are 0.2006 and 
420 km s$^{-1}$, respectively. The estimated velocity dispersion 
has a large confidence interval which is more likely to be a result of poor sampling.
Figure~\ref{fig:sub5coutour} shows the isoline plot of Structure 9 (blue dots with label 9) which is superimposed on
the scatter plot. Details of this structure and its member galaxies are given in Table~\ref{tab:sub2groups}.
\end{itemize}
%%%%%%%%%%%%%%%%%%%%%%%%%%%%%

%%%%%%%%%%%%%%%%%%%%%%%%%%%%%%%%%%%%%%%%%%%%%%

We note that there are four other galaxy clusters reported in the literature in a one degree radius around A3888 
(ABELL 3896 NED01 \& ABELL 3896 NED02, \citealt{1999ApJ...520..491B}, ABELL S1045, \citealt{1994MNRAS.269..151D} and
ABELL S1051, \citealt{1995MNRAS.274.1071C}). There are also three galaxy groups reported 
in this region (LCLG -39 198 \& LCLG -39 199 \citealt{2000ApJS..130..237T} and USGC S275, 
\citealt{2000ApJ...534..114M}) however, none of there were re-detected in this work.
Given the large distance of these clusters to A3888 and our poor redshift coverage in the far field of A3888, it was highly 
unlikely for us to find these clusters.

Based on an search for galaxy over-densities in the optical images associated with the LARCs project using the local galaxy density parameter, $\Sigma$, \citet{2001PhDT........74P}  identified six over-densities of which he put forward five candidate clusters in the field around A3888. Examination of colour magnitude diagrams suggested that several of these were likely clusters, however there was a lack of spectroscopic information for the majority of putative cluster members. Of note are three objects listed in \citet{2001PhDT........74P} which he denoted $\alpha$, $\beta$ and $\epsilon$ which correspond to Structures 7, 6 and 2, respectively. Comparison of the redshifts present by \citet{2001PhDT........74P} and those given here shows that for Structure 7 -- SJD J2232.5-3759 the single redshift presented by \citet{2001PhDT........74P} is one of the 24 cluster members confirmed here and for Structure 6 -- SJD J2232.4-3742 four of the five redshifts listed by \citet{2001PhDT........74P} are part of 
the 17 members listed here and in both cases the use of the local density parameter identifies the same cD galaxy as found here. In the case of Structure 2, five of the seven members were listed previously though we classify this as a group, not a cluster (see Section \ref{sec:sec7}). The three remaining putative over-densities listed in \citet{2001PhDT........74P} have not been detected in this work.

%%%%%%%%%%%%%%%%%%%%%%%%%%%%%%%%%%%%%%%%%%%%%%%%%%%%%%%%%%%%%%%%%%%%%%%

\section{Discussion and Conclusion}
\label{sec:sec10}
We have presented the results of new optical observations of the cluster A3888 taken with the
AAOmega spectrograph. We present 254 new redshifts in the region of A3888. 
Combining these with available redshifts in the literature we find
that main cluster, A3888, has 71 member galaxies, a redshift of 0.1535$\pm$0.0009 and a rest frame velocity dispersion of 1181$\pm$197 km/s. 
The cluster is elongated along an East-West axis and a 3D Lee-Fitchett substructure test of the cluster indicates that A3888 is a bimodal cluster along that same axis.

The contradictory evidence about the dynamics of A3888 from X-ray data explained earlier underlines that the dynamical 
status of this cluster was previously not clear. 
However, the combination of pieces of evidence from the optical analysis such as the presence of multiple BCGs, 
the elongated optical galaxy distribution, and our substructure test which showed that A3888 is bimodal 
strongly suggests that this cluster has had dynamical interactions and is highly likely to be a young 
cluster in an active merging state. This is consistent with the latest X-ray morphology reported by 
\citet{2013A&A...549A..19W} which confirmed that A3888 has local X-ray substructures and a large BCG/X-ray peak separation. 

Further spectroscopic analysis of this cluster would be useful to further probe its dynamics, however, on account of the very small
angular separation of the galaxies in the core of the cluster, single slit
spectroscopy or more usefully observations with an integral field unit are required to increase the spectroscopic coverage in the cluster core. 
%In particular, the 2.3~m telescope
%of the Australian National University (ANU) at Siding Spring Observatory with rotating slit could be used in the future to 
%cover two or more galaxies in a single exposure \citep{2006AJ....131.1280F} greatly reducing observing time required to cover the dense core region of the cluster. 
This would allow a more detailed probe of the cluster core and better statistics on the merging populations.

In addition to the work on A3888 we also presented 
six galaxy over-densities within a one degree radius of the clusters, 
including three new spectroscopically detected galaxy clusters. Again, further observations of these structures, 
particularly the clusters, would be beneficial.

\section*{Acknowledgements}
We thank the anonymous referee for the excellent suggestions which improved the quality of this work and for drawing our attention to the thesis of Pimbblet which was not available in ADS. We further wish to acknowledge the AAT staff for performing the AAOmega observations in service mode, and Dr Sarah Brough for her contribution to our AAOmega proposal. SS was funded in this work by a Victoria Doctoral Scholarship. This publication is based on data acquired through the Australian Astronomical Observatory, under program AO163. The Digitized Sky Surveys were produced at the Space Telescope Science Institute under U.S. Government grant NAG W-2166. This research has made use of the NASA/IPAC Extragalactic Database (NED) which is operated by the Jet Propulsion Laboratory, 
California Institute of Technology, under contract with the National Aeronautics and Space Administration.
%%%%%%%%%%%%%%%%%%%%%%%%%%%%%%%%%%%%%%%%%%%%%%%%%%

%%%%%%%%%%%%%%%%%%%% REFERENCES %%%%%%%%%%%%%%%%%%

\bibliographystyle{mnras}
\bibliography{A3888_FINAL}

%%%%%%%%%%%%%%%%% APPENDICES %%%%%%%%%%%%%%%%%%%%%

\appendix
 \section{Member Galaxies of the Newly Detected Galaxy Clusters}
\begin{table}
\caption{Details of the galaxies of detected galaxy clusters (Structures 6, 7, 8).}
 \begin{tabular}{ccc}
 \hline
 \\[-10pt]
 R.A&Dec&z\\
\\[-10pt]
  \hline
  \\[-10pt]
  \multicolumn{3}{c}{Structure 6 -- SJD J2232.4-3742}\\
  \\[-10pt]
  \hline
 22 32 22.70	&	 -37  44 54.00	&	0.07743	\\
 22 32 29.68	&	 -37  44 38.11	&	0.08022	\\
 22 32 31.70	&	 -37  42 01.00	&	0.07998	\\
 22 32 35.90	&	 -37  42 31.00	&	0.07945	\\
 22 32 37.40	&	 -37  41 23.00	&	0.07795	\\
 22 32 38.75	&	 -37  41 38.18	&	0.07834	\\
 22 32 38.89	&	 -37  46 52.21	&	0.08007	\\
 22 32 39.03	&	 -37  39 48.06	&	0.07909	\\
 22 32 40.12	&	 -37  42 22.18	&	0.07814	\\
 22 32 40.60	&	 -37  42 16.00	&	0.07790	\\
 22 32 41.00	&	 -37  43 10.00	&	0.08454	\\
 22 32 42.90	&	 -37  44 29.00	&	0.07972	\\
 22 32 43.70	&	 -37  48 13.00	&	0.08439	\\
 22 32 45.60	&	 -37  43 56.00	&	0.08297	\\
 22 32 49.00	&	 -37  41 47.00	&	0.08095	\\
 22 32 53.80	&	 -37  41 41.00	&	0.08433	\\
 22 32 56.60	&	 -37  44 40.00	&	0.07692	\\
\hline
\\[-10pt]
 \multicolumn{3}{c}{Structure 7 -- SJD J2232.5-3759}\\
  \\[-10pt]
  \hline

22 32 09.10	&	 -38 01 01.00	&	0.07401	\\
22 33 10.20	&	 -37 57 56.00	&	0.07366	\\
22 32 15.20	&	 -38 00 20.00	&	0.07345	\\
22 32 24.70	&	 -38 04 29.00	&	0.07361	\\
22 32 29.00	&	 -38 05 19.00	&	0.07449	\\
22 32 32.10	&	 -38 01 25.00	&	0.07926	\\
22 32 38.20	&	 -37 56 36.00	&	0.07813	\\
22 32 41.10	&	 -37 59 33.00	&	0.07614	\\
22 32 49.40	&	 -37 59 24.00	&	0.07413	\\
22 32 51.00	&	 -37 53 33.00	&	0.07906	\\
22 32 51.10	&	 -38 00 39.00	&	0.07411	\\
22 33 00.91	&	 -37 57 12.20	&	0.07203	\\
22 33 04.30	&	 -37 53 29.00	&	0.07828	\\
22 33 04.61	&	 -37 53 50.89	&	0.07865	\\
22 33 05.80	&	 -37 53 02.00	&	0.07881	\\
22 33 08.70	&	 -37 58 52.00	&	0.07410	\\
22 33 09.40	&	 -37 52 47.00	&	0.07859	\\
22 33 10.20	&	 -37 57 56.00	&	0.07366	\\
22 33 10.60	&	 -37 50 30.00	&	0.07834	\\
22 33 10.70	&	 -37 55 41.00	&	0.07847	\\
22 33 13.00	&	 -37 54 05.00	&	0.07891	\\ 
22 33 13.26	&	 -37 49 36.16	&	0.07845	\\ 
22 33 17.34	&	 -37 56 05.10	&	0.07970	\\ 
22 33 24.90	&	 -37 51 44.00	&	0.07473	\\ 
\hline
\\[-10pt]
 \multicolumn{3}{c}{Structure 8 -- SJD J2231.1-3801}\\
 \hline
  \\[-10pt]
 22 30 34.70	&	 -38 01 00.00	&	0.07436	\\
 22 30 44.00	&	 -38 03 31.00	&	0.07253	\\
 22 30 50.10	&	 -38 02 47.00	&	0.07351	\\ 
 22 30 53.90	&	 -38 00 20.00	&	0.07312	\\
 22 30 58.20	&	 -38 03 51.00	&	0.07327	\\
 22 31 07.20	&	 -38 01 36.00	&	0.07329	\\
 22 31 10.10	&	 -38 01 16.00	&	0.07325	\\
\\[-10pt]
\hline
 \end{tabular}
 \label{tab:new-clust}
\end{table}
\bsp
\label{lastpage}
\end{document}